\documentclass[12pt]{iopart}

\usepackage{citecollapse}

\usepackage{graphicx}
\usepackage{iopams}
\usepackage{color}

\newcommand{\tensor}[1]{\stackrel{\leftrightarrow}{#1}} 

\begin{document}

\title[Numerical electrokinetics]{Numerical electrokinetics}

\author{R. Schmitz$^1$ and B. D\"unweg$^{1,2}$}
\address{$^1$ Max Planck Institute for Polymer Research,
Ackermannweg 10, 55128 Mainz, Germany \\
$^2$ Department of Chemical Engineering, Monash University,
Clayton, Victoria 3800, Australia}

\begin{abstract}

  A new lattice method is presented in order to efficiently solve the
  electrokinetic equations, which describe the structure and dynamics
  of the charge cloud and the flow field surrounding a single charged
  colloidal sphere, or a fixed array of such objects. We focus on
  calculating the electrophoretic mobility in the limit of small
  driving field, and systematically linearise the equations with
  respect to the latter. This gives rise to several subproblems, each
  of which is solved by a specialised numerical algorithm. For the
  total problem we combine these solvers in an iterative
  procedure. Applying this method, we study the effect of the
  screening mechanism (salt screening vs. counterion screening) on the
  electrophoretic mobility, and find a weak non--trivial dependence, as
  expected from scaling theory. Furthermore, we find that the
  orientation of the charge cloud (i.~e. its dipole moment) depends on
  the value of the colloid charge, as a result of a competition
  between electrostatic and hydrodynamic effects.

\end{abstract}

\pacs{
  47.11.-j,     
  47.57.jd,     
  47.57.E-,     
  47.57.J-,     
  47.57.-s,     
  82.45.-h,     
  82.70.Dd,     
  82.70.-y,     
  05.60.Cd      
}

\maketitle

\section{Introduction}
\label{sec:intro}

  The interplay between electrostatic and hydrodynamic interactions
  is of high importance for the understanding of a wide range of
  biological, chemical and physical systems, since in almost all
  situations where a solid is brought in contact with a liquid,
  a difference in the electric potential occurs due to association
  or dissociation of charges or the orientation of molecules at
  the surface.

  Charged solid colloidal spheres in suspension in an aqueous solution
  containing counterions and salt ions will be surrounded by a cloud
  of oppositely charged ions. This cloud, typically called the
  electric double layer, is responsible for screening the
  electrostatic potential. If an external electric field is acting on
  the system, the charged spheres start to migrate in the direction of
  the oppositely charged electrode and the surrounding cloud will be
  deformed and becomes anisotropic due to the electric field and also
  to the friction between the ions and the fluid. This phenomenon is
  called electrophoresis and the corresponding transport coefficient
  is the electrophoretic mobility $\mu$, determined by the balance of
  electric driving and hydrodynamic frictional force acting on the
  sphere. It is defined as the proportionality constant between the
  constant velocity $\boldsymbol{u}$ of the particle and the external
  driving field $\boldsymbol{E}_{ext}$ in the linear regime, i.~e. for
  small driving fields,
  \begin{equation} \label{eq: electrophoretic mobility}
      \boldsymbol{u} = \mu \boldsymbol{E}_{ext} \,.
  \end{equation}
  Efforts have been made to study electrophoresis by experimental
  methods \cite{Eve98,Med2003,Med2004,Gar2004,Pal2004} as well as by
  analytical and numerical calculations over the decades
  \cite{Smo03,Hue24,Hen31,Ove43,Boo50,Wie66,Ohs83,OBr78,Man92,%
    lozada1999nonlinear,lozada2001primitive,Ohs2000,%
    Mis2002,Fre2004,Kim06,Lobaskin2004,Lobaskin2007,Due08,All2010}.
  Due to the complicated many--body nature of the problem a
  comprehensive quantitative theoretical understanding is still
  lacking.

  An important milestone was the theoretical investigation by O'Brien
  and White \cite{OBr78}, later known as ``standard electrokinetic
  model''. Starting from the dynamic Mean--Field equations that
  describe the interplay between the convection--diffusion dynamics of
  the ion clouds, the solvent flow field, and the electrostatic forces
  in the system, they studied a single charged colloidal sphere in
  infinite space, subject to an infinitesimally weak homogeneous
  external electric field, with respect to which the equations are
  linearised. This problem exhibits full spherical symmetry, and hence
  its numerical treatment reduces to the solution of a
  one--dimensional ordinary differential equation, which finally
  allows the calculation of $\mu$.

  This approach has a fairly broad but not unlimited range of
  applicability, whose conditions may be summarised as follows:
  Firstly, the Mean--Field theory as such must be justified, and this
  implies weak ion--ion correlations, which is typically the case for
  single--valence ions at room temperature. Furthermore, treating the
  problem within the framework of a single--colloid theory requires
  that the ionic clouds essentially do not overlap, and all
  non--trivial values of charge density, electrostatic potential, and
  flow velocity are confined to the double layer, as a result of
  electrostatic and hydrodynamic screening. This means that the salt
  concentration has to be fairly large, and the electrostatic
  screening is dominated by the salt contribution: Note that for a
  single sphere in infinite space the counterions have all
  entropically ``evaporated'', and hence must be ignored in the
  theory.
  
  It is exactly this latter condition that is violated in recent
  experiments on colloidal electrophoresis \cite{Eve98,Lobaskin2007},
  which have deliberately focused on the limit of low salt. In
  Ref. \cite{Lobaskin2007} it was shown that some (certainly not all)
  experimental observations in that regime can be explained by
  assuming that the effects of finite colloid volume fraction (and
  corresponding finite counterion concentration) can be modeled by
  simply studying a single colloidal particle in a finite simulation
  box with periodic boundary conditions, which automatically gives
  rise to a finite colloid volume fraction and the correct
  corresponding counterion concentration. This investigation was done
  by studying a system of charges by Molecular Dynamics, while the
  solvent hydrodynamics was taken into account by a Lattice Boltzmann
  background. However, the computational effort of such studies turned
  out to be large, such that it was neither possible to obtain highly
  accurate results, nor to vary parameters systematically over a broad
  range.

  For these reasons, we develop a new approach in the present paper:
  On the one hand, we wish to study precisely the same physical
  situation as in Ref. \cite{Lobaskin2007}, i.~e. a single colloidal
  particle in a finite box with a finite counterion concentration,
  possibly with added salt, while on the other hand taking full
  advantage of the O'Brien and White approach, which means that we
  study the Mean--Field equations, combined with linearisation with
  respect to the driving field. In summary, our work is nothing more
  and nothing less than the finite--volume generalisation of
  Ref. \cite{OBr78}, whose results are directly recovered in the
  special limit of large salt concentration, where the double layer is
  significantly smaller than the box. Our method is based upon a full
  three--dimensional calculation, where the partial differential
  equations are discretised on a lattice. Since these equations are
  mathematically fairly different, we develop specialised solvers for
  each equation. The colloidal particle acts essentially as a boundary
  condition, and hence it is clear that the method can be easily
  generalised to a fixed array of particles or any other object that
  is periodically repeated in space --- strictly spoken, we are
  studying a colloidal crystal, as a result of the periodic boundary
  conditions of the box. Within this setup one may vary several
  important parameters like volume fraction and averaged ion
  concentrations, and study the behaviour of $\mu$ fairly accurately.

  Returning to the issue of the limitations of the present approach,
  we notice that the present model differs from the Molecular
  Dynamics~/ Lattice Boltzmann (MD/LB) model not only with respect to
  its computational cost, but also in terms of the modeling of the
  finite size of the ions. The present model clearly assumes point
  ions and neglects any ion--ion correlations, which, however, due to
  packing effects, are important if the colloid size is not very large
  compared to the ion size
  \cite{lozada1999nonlinear,lozada2001primitive}. On the other hand,
  these correlations are fully present in the MD/LB model. However, it
  should be noted that the MD/LB model is, for computational reasons,
  very limited in terms of the ratio of colloid size to ion size,
  which cannot have values much in excess of $10$. Therefore, the
  MD/LB model will probably, in comparison to experiment, overestimate
  the effects of packing and ion correlations, except for quite small
  colloids. We therefore study here the other extreme, which should be
  reasonable for large colloids. While there are attempts to include
  the finite size of ions into more generalised Mean Field theories,
  both for statics \cite{lozada1999nonlinear,lozada2001primitive,%
    Li2011} and dynamics
  \cite{lozada1999nonlinear,lozada2001primitive}, it is not
  immediately obvious how to incorporate these formalisms into a
  three--dimensional code that is strictly confined to a finite box
  with a well--defined and conserved number of ions. Furthermore, the
  present model should be viewed as just a first step to the
  development of a more general simulation program that is able to
  study multi--colloid systems. At finite volume fraction,
  colloid--colloid correlations are quite important, and they are of
  course not taken into account by our ``cell'' or ``colloidal
  crystal'' model. For the statics, the importance of such
  correlations has been pointed out, e.~g., in Refs.
  \cite{manzanilla2011zeta,manzanilla2011polarity}; in the dynamic
  case one expects a probably even stronger effect, since here not
  only the electrostatic interactions are insufficiently screened, but
  the hydrodynamic interactions as well. A multi--colloid simulation
  model with explicit ions is however very likely to be
  computationally too expensive, except for very moderate scale
  separations between macro- and micro-ions, both with respect to
  length and with respect to charge. The present work is intended as a
  first step in an attempt to overcome at least the former of these
  limitations, by assuming full scale separation between length scales
  at the outset. Of course, numerically this scale separation is
  anything but perfect, due to limited grid resolution; however,
  discretised field theories seem to be more amenable to extrapolation
  procedures than particle models.

  In Sec. \ref{sec:theory} the theoretical model is introduced,
  including the Mean--Field approach and the linearisation of the
  equations. Furthermore, the equations are reformulated in
  dimensionless units. The computational method is briefly discussed
  in Sec. \ref{sec:Comp}, where we describe the iterative combination
  of the specialised solvers and discuss the particular choices for
  the numerical methods. In Sec. \ref{sec:results} some interesting
  results from this method are presented: The dependence of the
  electrophoretic mobility on the details of the screening mechanism
  is analysed, and the reversal of the field--induced dipole moment of
  the ion cloud surrounding a weakly charged colloid is elucidated.
  Section \ref{sec:conclusions} concludes with a brief summary.

  \section{Theory}
  \label{sec:theory}

  \subsection{Electrokinetic equations}
  \label{ssec:SEM}

  Electrophoresis is a result of the balance between electrostatics
  and hydrodynamics. Within a Mean--Field picture the system is
  described in terms of ion concentration fields $c_i$, electrostatic
  potential $\psi$ and the flow velocity field
  $\boldsymbol{v}$. Cross--correlations between salt ions as well as
  thermal fluctuations are neglected.

  The Poisson equation couples the concentration fields to the
  electrostatic potential,
  \begin{equation} \label{eq: Poisson}
      - \nabla^2 \psi = \frac{1}{\varepsilon} e \sum_i z_i c_i\,.
  \end{equation}
  Here, $\varepsilon$ is the dielectric constant, $e$ denotes the
  elementary charge and $z_i$ is the valence of the ionic species,
  where the subscript $i$ indicates the different ionic species in
  the system. Counterions, which assure the charge neutrality of
  the system, are denoted by the index $i=0$. The charged colloids
  are taken into account via boundary conditions.

  The dynamics of the concentration field $c_i$ is described by a
  continuity equation, where the total current density is a
  combination of a diffusive term, a convective current and the
  current resulting from the electric force. One thus obtains a
  convection--diffusion equation, known as Nernst--Planck equation,
  \begin{equation} \label{eq:NernstPlanck}
    \partial_t c_i = \nabla \cdot \left(
            D_i \nabla c_i
            + \frac{D_i}{k_B T} e z_i (\nabla \psi) c_i
            - \boldsymbol{v} c_i
            \right) \,. 
  \end{equation}
  Here, $D_i$ is the diffusion constant of the ionic species $i$ and
  $k_B T$ denotes the thermal energy. The ion mobility is given by
  $D_i / (k_B T)$ due to the Einstein relation. 

  Electric forces and viscous forces are balanced in the Stokes
  equation which describes zero Reynolds number incompressible
  hydrodynamics,
  \begin{eqnarray} 
      \label{eq: incompressibility}
      \nabla \cdot \boldsymbol{v} & = & 0 \, , \\
      \label{eq: Stokes}
      \rho \partial_t \boldsymbol{v} & = &
      - \nabla p + \eta \nabla^2 \boldsymbol{v}
      - e (\nabla \psi) \sum_i z_i c_i \,,
  \end{eqnarray}
  where $\rho$ is the mass density of the fluid, $p$ its pressure
  field and $\eta$ the fluid viscosity. 

  In the stationary state, the system of equations is thus summarised
  as \cite{Rus89}
  \begin{eqnarray}
      \label{eq:EKE_a}
      0 & = &
	  \nabla^2 \psi + \frac{1}{\varepsilon} e \sum_i z_i c_i \,,\\
      \label{eq:EKE_b}
      0 & = &
	  \nabla \cdot \left( D_i \nabla c_i
	  + \frac{D_i}{k_B T} e z_i (\nabla \psi) c_i
	  - \boldsymbol{v} c_i \right) \,, \\
      \label{eq:EKE_c}
      0 & = &
	  - \nabla p + \eta \nabla^2 \boldsymbol{v}
	  - e (\nabla \psi) \sum_i z_i c_i\,,\\
      \label{eq:EKE_d}
      0 & = &
	  \nabla \cdot \boldsymbol{v}\,.
  \end{eqnarray}

  Notice that the stationary formulation is not manifestly Galilei
  invariant, but rather selects one particular frame of reference (the
  rest frame of the colloidal particle) in which it is valid. If a
  colloidal sphere would move relative to the chosen inertial frame,
  the local ionic concentration would change with time, and hence a
  stationary solution would not exist. It is this restriction which
  either confines the method to single--colloid studies, or forces us
  to impose a somewhat unphysical ``rigid-body'' constraint between
  the set of colloidal particles. The mobility however is measured in
  the system's center--of--mass reference frame. In other words, the
  center--of--mass velocity of the system must be taken as the
  velocity that determines $\mu$. As a matter of fact, it turned out
  that it is most convenient to solve the Nernst--Planck equation in
  the colloid rest frame, while the Stokes equation is best solved in
  the rest frame of the center of mass. Therefore, one always needs a
  trivial Galilei transform when switching from one equation to the
  other.

  \subsection{Dimensionless formulation}
  \label{ssec:dimless}

  An important length scale in the theory of charged systems is the
  Bjerrum length, which results from the balance between electrostatic
  and thermal energy:
  \begin{equation} \label{eq:Bjerrum_length}
      l_B = \frac{e^2}{4 \pi \varepsilon k_B T} \,.
  \end{equation}

  The Stokes mobility of a sphere of radius $l_B$ and elementary
  charge $e$ provides a natural unit for the electrophoretic mobility:
  \begin{equation} \label{eq:mu_0}
     \mu_0 = \frac{e}{6 \pi \eta l_B} \,,
  \end{equation}
  and the dimensionless reduced electrophoretic mobility is
  defined as
  \begin{equation} \label{eq:reduced_mobility}
      \mu_{red} = \frac{\mu}{\mu_0} \,.
  \end{equation}
  The natural energy scale is the thermal energy $k_B T$,
  and together with the elementary charge $e$ this yields
  a dimensionless electrostatic potential
  \begin{equation} \label{eq:scale_potential}
      \tilde \psi = \psi \frac{e}{k_B T}\,.
  \end{equation}
  Introducing a second length scale $\kappa^{-1}$ (see below), such
  that the gradient is rescaled via
  \begin{equation} \label{eq:Def_scale_parameter}
    \tilde \nabla \, = \, \frac{1}{\kappa} \nabla \,,
  \end{equation}
  the electrokinetic equations are nondimensionalised as
  \begin{eqnarray}
      0 & = &
	  {\tilde \nabla}^2 \tilde \psi + \sum_i z_i {\tilde c}_i \,,\\
      0 & = &
	  \tilde \nabla \cdot \left( {\tilde D}_i \tilde \nabla {\tilde c}_i
	  + {\tilde D}_i z_i (\tilde \nabla \tilde \psi) {\tilde c}_i
	  - \tilde{\boldsymbol{v}} {\tilde c}_i \right) \,, \\
      0 & = &
	  - \tilde \nabla \tilde p + \frac{2}{3} {\tilde \nabla}^2
	    \tilde{\boldsymbol{v}}
	  - (\tilde \nabla \tilde \psi) \sum_i z_i {\tilde c}_i\,,\\
      0 & = &
	  \tilde \nabla \cdot \tilde{\boldsymbol{v}}\, ,
  \end{eqnarray}
  where the dimensionless parameters and variables are summarised in
  Tab. \ref{Tab:Reduced Parameters}.
  \begin{table*}
  \noindent
  \begin{centering}
    \begin{tabular}{l||c c}
      parameter & physical units & dimensionless formulation 
      \tabularnewline
      \hline
      \hline
      \tabularnewline
      Bjerrum length & $l_B = \frac{e^2}{4\pi \varepsilon k_B T}$ &
      \tabularnewline
      \tabularnewline
      screening parameter & $\kappa^2 = 4\pi l_B \sum_i z_i^2 \frac{N_i}{V}$ &
      \tabularnewline
      \tabularnewline
      electrophoretic mobility & $\mu$ & 
             $\mu_{red} = \frac{6 \pi \eta l_B}{e} \mu$
      \tabularnewline
      \tabularnewline
      spatial position & $\boldsymbol{r}$ &
            $\tilde{\boldsymbol{r}} = \kappa \boldsymbol{r}$
      \tabularnewline
      \tabularnewline
      spatial derivative & $\nabla$ & 
            $\tilde \nabla = \frac{1}{\kappa} \nabla$
      \tabularnewline
      \tabularnewline
      electrostatic potential & $\psi$ & 
            $\tilde \psi = \frac{e}{k_B T} \psi$
      \tabularnewline
      \tabularnewline
      electric field& $\boldsymbol{E}$ &
            $\tilde{\boldsymbol{E}} = \frac{e}{\kappa k_B T}\boldsymbol{E}$
      \tabularnewline
      \tabularnewline
      ion concentration & $c_i$ & 
            ${\tilde c}_i = \frac{4\pi l_B}{\kappa^2} c_i$
      \tabularnewline
      \tabularnewline
      colloid charge & $Ze$ &
            $\tilde Z = 4\pi l_B \kappa Z$
      \tabularnewline
      \tabularnewline
      number of ions & $N_i$ & 
            ${\tilde N}_i = 4\pi l_B \kappa N_i$
      \tabularnewline
      \tabularnewline
      flow velocity & $\boldsymbol{v}$ &
            $\tilde{\boldsymbol{v}} = 
            \frac{6\pi l_B \eta}{\kappa k_B T} \boldsymbol{v}$
      \tabularnewline
      \tabularnewline
      pressure & $p$ & 
            $\tilde p = \frac{4\pi l_B}{\kappa^2 k_B T} p$
      \tabularnewline
      \tabularnewline
      diffusion constant & $D_i$ &
            ${\tilde D}_i = \frac{6\pi l_B \eta}{k_B T} D_i$
      \tabularnewline
    \end{tabular}
    \par
    \end{centering}
    \caption{Summary of all parameters in physical units and their
	    reduced counterparts.} 
    \label{Tab:Reduced Parameters}
  \end{table*}

  Note that the transformation from physical to reduced units, as
  outlined in Tab. \ref{Tab:Reduced Parameters}, is valid for any
  arbitrary choice of the length scale $\kappa^{-1}$. However, a
  physically motivated choice results from the finite--volume version
  of linearised Poisson--Boltzmann theory (Debye-H\"uckel theory). We
  therefore choose $\kappa$ to be the Debye screening parameter,
  \begin{equation}\label{eq:Debye_screening}
    \kappa^2 = 4 \pi l_B \sum_i z_i^2 \frac{N_i}{V} \,,
  \end{equation}
  where all ionic species (including the counterions) contribute, $V$
  is the volume of the system (actually the volume that is available
  to the ions, i.~e. box volume minus colloid volume), and $N_i$ the
  number of ions of species $i$. For simplicity the tilde will from
  now on be omitted, with the understanding that all parameters are
  given in reduced units.

  \subsection{Linearisation}
  \label{ssec:linearization}

  The high nonlinearity of the Mean--Field equations causes two
  problems. Firstly, it is difficult and computationally expensive to
  solve a coupled system of nonlinear differential
  equations. Furthermore, the electrophoretic mobility is
  well--defined (i.~e. independent of the driving field) only in the
  linear regime. Consequently, if a fully nonlinear solution of the
  equations is obtained, an extrapolation to zero driving field is
  required. The second problem can be avoided completely, and the
  first one at least reduced, by a linearisation of the equations in
  terms of the driving field \cite{OBr78,Rus89}. This can be done by a
  formal expansion with respect to a small parameter $\epsilon$,
  corresponding to the strength of the external field. All fields in
  the system have a regular expansion in $\epsilon$, and hence may be
  written as
  \begin{eqnarray}
      \label{eq:exp1}
      c_i & = & c_i^{(0)} + 
      \epsilon c_i^{(1)} + \mathcal{O}(\epsilon^2) \,, \\
      \label{eq:exp2}
      \psi & = & \psi^{(0)} +
      \epsilon \psi^{(1)} + \mathcal{O}(\epsilon^2)\,, \\
      \label{eq:exp3}
      {\boldsymbol{v}} & = & \epsilon {\boldsymbol{v}}^{(1)} 
      + \mathcal{O}(\epsilon^2)\,,\\
      \label{eq:exp4}
      p & = & p^{(0)} + \epsilon p^{(1)} + 
      \mathcal{O}(\epsilon^2) \, ;
  \end{eqnarray}
  note that for $\epsilon = 0$, i.~e. in the absence of external
  driving, the system is at rest, such that the zeroth--order velocity
  vanishes. We now insert the expansion into the electrokinetic equations;
  noticing that for $\epsilon = 0$ all ionic currents must vanish,
  one obtains
  \begin{itemize}
  \item  in zeroth order perturbation theory:
	  \begin{eqnarray} 
	      0 & = & \nabla^2 \psi^{(0)} + \sum_i z_i c_i^{(0)} \,,
	      \label{eq: zeroth order EKE 1}\\
	      0 & = & \nabla \left( z_i \psi^{(0)} + \ln c_i^{(0)} \right) \,,
	      \label{eq: zeroth order EKE 2}\\
	      0 & = & - \nabla p^{(0)} - (\nabla \psi^{(0)}) 
                      \sum_i z_i c_i^{(0)}\,.
	      \label{eq: zeroth order EKE 3}
	  \end{eqnarray}

  \item  and in first order:
	  \begin{eqnarray}
	      0 & = & \nabla \cdot \bigg\{ D_i \nabla c_i^{(1)}
		      + D_i z_i (\nabla \psi^{(1)}) c_i^{(0)} \nonumber \\
		 &&   + D_i z_i (\nabla \psi^{(0)}) c_i^{(1)}
		      - {\boldsymbol{v}}^{(1)} c_i^{(0)} \bigg\}\,,
		      \label{eq: first order EKE a}\\
	      0 & = & - \nabla p^{(1)} + \frac{2}{3} \nabla^2
		      {\boldsymbol{v}}^{(1)} 
		      - (\nabla \psi^{(1)}) \sum_i z_i c_i^{(0)}
		      - (\nabla \psi^{(0)}) \sum_i z_i c_i^{(1)}\,,
		      \label{eq: first order EKE b}\\
	      0 & = & \nabla \cdot {\boldsymbol{v}}^{(1)}\,,
		      \label{eq: first order EKE c}\\
	      0 & = & \nabla^2 \psi^{(1)} + \sum_i z_i c_i^{(1)}\,.
		      \label{eq: first order EKE d}
	  \end{eqnarray}

  \end{itemize}

  The zeroth order only contains the electrostatic potential of the
  unperturbed ion clouds; hence this order is identical to standard
  nonlinear Poisson--Boltzmann theory. Equation \ref{eq: zeroth order
    EKE 3} is just an equation to determine the zeroth--order
  pressure, which is of no interest to us; it can therefore be simply
  ignored. The first order consists of a coupled set of \emph{linear}
  equations; hence the only nonlinearity that remains is the
  equilibrium Poisson--Boltzmann problem, which is simpler than
  studying the original full set of nonlinear dynamic equations.

  In the first--order equations, the external field is taken into
  account by decomposing the potential $\psi^{(1)}$ into a periodic
  part and one part corresponding to the constant electric field
  \begin{equation}
      \psi^{(1)} \, = \, \psi'^{(1)} + \psi''^{(1)} \,,
  \end{equation}
  such that
  \begin{eqnarray}
      \nabla^2 \psi'^{(1)} & = & - \sum_{i} z_i c_i^{(1)} \,,\\
      \nabla^2 \psi''^{(1)} & = & 0 \,,\\
      \nabla \psi''^{(1)} & = & - \boldsymbol{E}_{ext} \,.
  \end{eqnarray}
  Hence one may write the first--order equations more explicitly as
  \begin{eqnarray} \label{eq: first order EKE 2}
      0 & = & \nabla \cdot \bigg\{ D_i \nabla c_i^{(1)}
	      + D_i z_i (\nabla \psi'^{(1)}) c_i^{(0)}
	      - D_i z_i \boldsymbol{E}_{ext} c_i^{(0)}\nonumber\\
	    && + D_i z_i (\nabla \psi^{(0)}) c_i^{(1)}
	      - {\boldsymbol{v}}^{(1)} c_i^{(0)} \bigg\} \,,
	      \label{eq: first order EKE 2 a}\\
      0 & = & - \nabla p^{(1)} + \frac{2}{3} \nabla^2 {\boldsymbol{v}}^{(1)}
	      - (\nabla \psi'^{(1)}) \sum_i z_i c_i^{(0)} \nonumber\\
	    && + \boldsymbol{E}_{ext} \sum_i z_i c_i^{(0)}
	      - (\nabla \psi^{(0)}) \sum_i z_i c_i^{(1)} \,,
	      \label{eq: first order EKE 2 b}\\
      0 & = & \nabla \cdot {\boldsymbol{v}}^{(1)} \,,
	      \label{eq: first order EKE 2 c}\\
      0 & = & \nabla^2 \psi'^{(1)} + \sum_i z_i c_i^{(1)} \,.
	      \label{eq: first order EKE 2 d}
  \end{eqnarray}
  It should be noted that, as a result of the perturbation expansion,
  the first--order convection--diffusion equation now contains sources
  and sinks (the terms proportional to $c_i^{(0)}$). However, since
  these terms all have the form of a divergence, there is neither a
  total flux of matter into the system, nor out of it, as it should
  be, since mass conservation must hold at each order of the expansion
  separately.

  The reduced electrophoretic mobility is then finally calculated as
  \begin{equation} \label{eq: electrophoretic mobility by expansion}
    \mu_{red} = \frac{\left\vert {\boldsymbol{u}}^{(1)} \right\vert}
    {\left\vert {\boldsymbol{E}}_{ext} \right\vert} \,,
  \end{equation}
  where ${\boldsymbol{u}}^{(1)}$ is the constant velocity of the
  colloid in the system's center--of--mass reference frame, i.~e.
  (assuming no--slip boundary conditions) the value of the flow
  velocity field at the surface of the colloidal sphere,
  \begin{equation}
     {\boldsymbol{u}}^{(1)} = {\boldsymbol{v}}^{(1)} (R) \,,
  \end{equation}
  with $R$ the radius of the particle. This mobility is strictly
  independent of the strength of the external driving field.

  \begin{figure}
    \begin{center}
      \includegraphics[clip,width=0.8\textwidth]
      {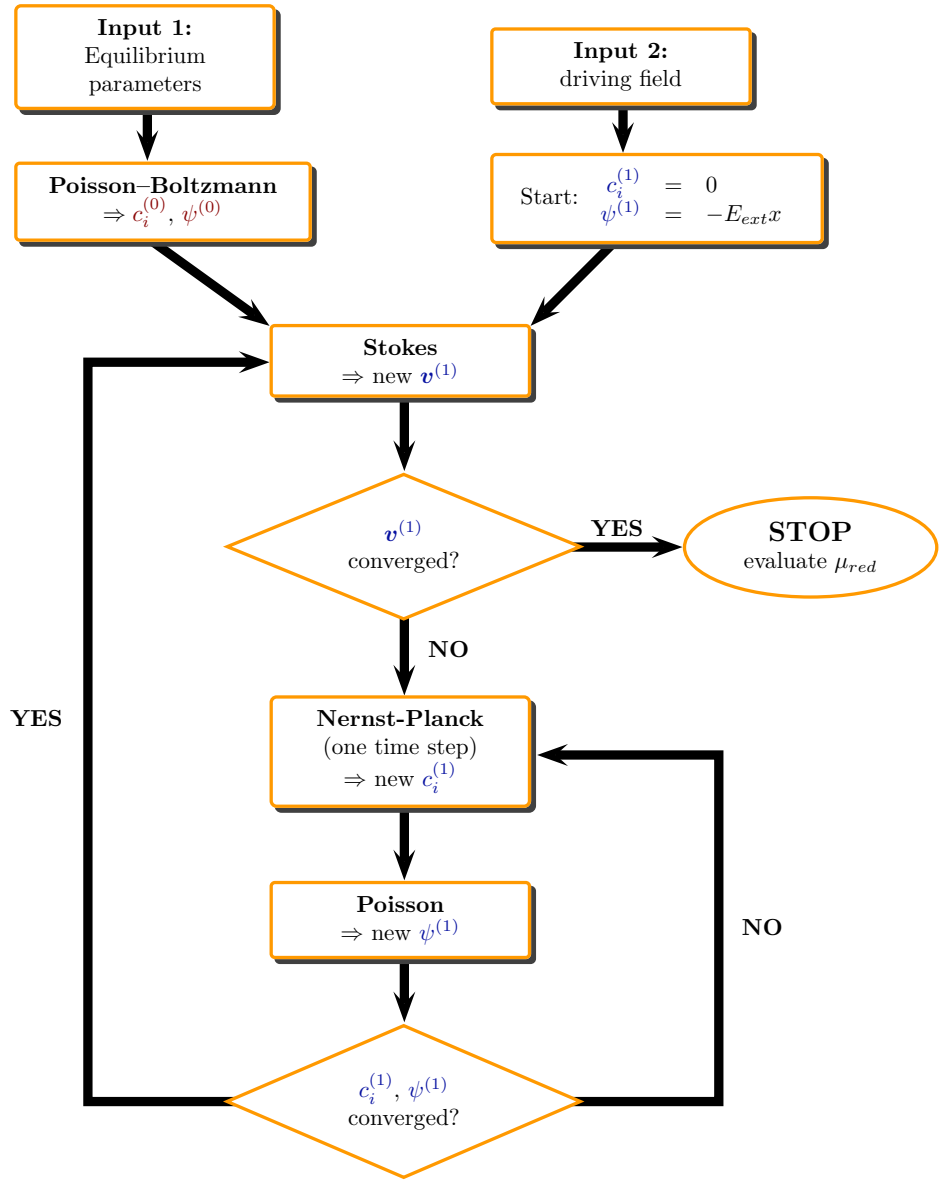}
    \end{center}
    \caption{Schematic illustration of the iterative algorithm for
             solving the electrokinetic equations.}
    \label{fig:IterativeProcedure}
  \end{figure}

  \section{Computational method}
  \label{sec:Comp}

  \subsection{Iterative procedure}
  \label{ssec:IterProc}

  The linearisation of the problem divides the challenge of solving
  the electrokinetic equations into two different subproblems.  For
  the zeroth order, a solution of the fully nonlinear
  Poisson--Boltzmann equation must be found. The first order consists
  of a set of linear equations, where the zeroth order fields only
  occur as prefactors. Here we need to solve the Poisson equation, the
  convection--diffusion equation and the incompressible Stokes
  equation. Each particular problem can then be dealt with by using a
  specialised solver, which is some finite--difference scheme on a
  regular lattice with periodic boundary conditions; details for each
  solver will be outlined below. Finally all methods are combined via
  iterative loops, as sketched in Fig. \ref{fig:IterativeProcedure}.

  It turned out that the convergence of the method is improved by not
  using the full new velocity field for the next iteration, but rather
  a convex combination of the result of the previous iteration and the
  most recent result of the Stokes solver $\boldsymbol{v}^{*}$:
  \begin{equation}
      \boldsymbol{v}^{(1)} \leftarrow 
      \omega \boldsymbol{v}^{*} + (1 - \omega) 
      \boldsymbol{v}^{(1)} \, , 
      \quad 0 < \omega \leq 1 \, ;
  \end{equation}
  in practice we used $\omega = 1/2$. The iteration procedure yields a
  sequence of reduced mobilities $\mu_{red}^{(i)}$, $i = 1, 2,
  \ldots$; the iteration was terminated as soon as the relative
  residual
  \begin{equation}
      \varepsilon \, = \, \left\vert \frac{\mu_{red}^{(i)}
      - \mu_{red}^{(i-1)}}{\mu_{red}^{(i)}} \right\vert
  \end{equation}
  dropped below the value $10^{-5}$.

  \subsection{Poisson-Boltzmann equation}
  \label{ssec:PB}

  For the zeroth order a solution of the fully nonlinear
  Poisson--Boltzmann equation is required. Here, a simple and
  unconditionally stable lattice algorithm has been used, based on a
  constrained variational approach of the Poisson--Boltzmann
  equation. This method has been discussed in detail in
  Ref. \cite{PB_PRE2009} and is hence only be briefly summarised
  here. It should be noted that this method has prompted other groups
  to develop similar ideas even further \cite{Li2011}; however, such
  more recent implementations have not been used here.

  Following the ideas of Maggs and Rosetto \cite{Maggs2002} and
  re--formulating the equations in terms of the electric field instead
  of the electrostatic potential, Eqs. \ref{eq: zeroth order EKE 1}
  and \ref{eq: zeroth order EKE 2} are written as
  \begin{eqnarray}
      \label{eq: Gauss law}
      \nabla \cdot \boldsymbol{E} & = & \sum_i z_i c_i ,\\
      \label{eq: curl E}
      \nabla \times \boldsymbol{E} & = & 0 , \\
      z_i \boldsymbol{E} & = & \nabla \ln c_i .
  \end{eqnarray}
  These equations are recovered as the Euler--Lagrange equations
  of a constrained free energy functional of the form
  \begin{eqnarray} \label{eq: functional}
      {\cal F} & = & \int_V f\, dV, \\
      f & = & \frac{1}{2} \boldsymbol{E}^2
      + \sum_i c_i \ln c_i
      - \psi ( \nabla \cdot \boldsymbol{E} -  \sum_i z_i c_i)
      - \sum_i \mu_i (c_i - \frac{N_i}{V}),
  \end{eqnarray}
  where the electrostatic potential $\psi$ and the chemical potential
  $\mu_i$ of ionic species $i$ occur as Lagrange multipliers taking
  into account Gauss' law and mass conservation, respectively. $N_i$
  is the total number of ions of the species $i$ and $V$ denotes the
  system's volume. Note that the formulation in terms of the electric
  field assures that the solution of the Poisson--Boltzmann equation
  is a true minimum of the functional. Applying a Yee discretisation
  \cite{Yee66}, i.~e. associating scalar fields with the sites, polar
  vectors with the links, and axial vectors with the plaquettes of a
  simple--cubic lattice, the functional can be minimised making local
  moves between adjacent nodes along a link and local field updates on
  the plaquettes. If the system has been initialised such that the
  constraints are fulfilled, those local moves never leave the
  constraint surface. Moreover, the update rules can be optimised,
  such that the functional value is decreased in every iterative step,
  and the method will run ultimately into the one and only minimum.
  For further details the reader is referred to
  Ref. \cite{PB_PRE2009}.

  \subsection{Poisson equation}
  \label{ssec:Poisson}

  The Poisson equation for a given charge density can be solved
  efficiently by Fast Fourier Transform. We expand the potential and
  the charge density in terms of Fourier series via
  \begin{eqnarray}
	\psi (\boldsymbol{r}) & = & \sum_{\boldsymbol{k}}
	\hat \psi (\boldsymbol{k}) 
        \exp [- i \boldsymbol{k} \cdot \boldsymbol{r}]\,, \\
	\rho (\boldsymbol{r}) & = & \sum_{\boldsymbol{k}}
	\hat \rho (\boldsymbol{k}) 
        \exp [- i \boldsymbol{k} \cdot \boldsymbol{r}]\,,
  \end{eqnarray}
  with
  \begin{equation}
        \boldsymbol{k} = 2 \pi (k / L_x, l/ L_y, m / L_Z) \, ,
	\quad k,l,m \in \mathbb{Z}\, .
  \end{equation}
  Here, $L_x \times L_y \times L_z$ is the dimension of the
  computational domain.
  The solution of the Poisson equation in Fourier space is
  then given by
  \begin{equation} \label{eq: Fourier transformed Poisson}
    \hat \psi = \frac{1}{\boldsymbol{k}^2} \hat \rho .
  \end{equation}

  For consistency reasons we use a discretised version, i.~e.  a
  lattice Green's function, instead of the continuum Green's function
  \cite{NumRec}; the discretised counterpart to Eq. \ref{eq: Fourier
    transformed Poisson} reads
  \begin{equation} \label{eq:DiscretePoisson_1}
      \hat \psi (k,l,m) = \frac{1}{\xi^2_{k,l,m}} 
      \hat \rho (k,l,m)
  \end{equation}
  with
  \begin{equation} \label{eq:DiscretePoisson_2}
      \xi_{k,l,m}^2  = \frac{2}{a^2} \left\{3 -
	\cos\left(2\pi \frac{k}{N_x}\right)
      - \cos\left(2\pi \frac{l}{N_y}\right)
      - \cos\left(2\pi \frac{m}{N_z}\right) \right\} \, ,
  \end{equation}
  where $a$ denotes the lattice spacing and $N_x = L_x/ a$ etc.
  Back--transformation finally yields the desired electrostatic potential
  in real space.

  \subsection{Stokes equation}
  \label{ssec:Stokes}

  The stationary incompressible Stokes equation has the form
  \begin{eqnarray} 
      \nabla \cdot \boldsymbol{v} (\boldsymbol{r}) & = & 0 \,,\\
      -\nabla p (\boldsymbol{r}) + \eta \nabla^2 \boldsymbol{v}
      (\boldsymbol{r}) & = & - \boldsymbol{f}_{ext} (\boldsymbol{r})\,,
  \end{eqnarray}
  where $\boldsymbol{f}_{ext}$ is an external force density and $\eta$
  denotes the fluid viscosity (which takes the value $2/3$ in our
  reduced unit system). For the purposes of the present paper, one
  should view $\boldsymbol{f}_{ext}$ as the force density generated by
  the electric field and the charges. More precisely, we include in
  $\boldsymbol{f}_{ext}$ all electric forces that come from the ion
  clouds, but also the force density that is generated from the fixed
  charges of the immersed body (or bodies). Since the total system is
  charge neutral, the total force on the system vanishes, even in the
  presence of external driving:
  \begin{equation}
     \int_V d^3 r \, \boldsymbol{f}_{ext} (\boldsymbol{r}) = 0.
  \end{equation}
  Again, it is obvious that this statement holds for the full
  nonlinear theory, and this implies that it must also hold separately
  at each order of the perturbation expansion. We now assume that the
  immersed bodies are not moving relatively to each other (cf. the
  remark at the end of Sec. \ref{ssec:SEM}), and also not
  rotating. Under these circumstances, the flow field can be
  calculated rather straightforwardly, making use of the idea of
  replacing the differential equations by equivalent integral
  equations \cite{You72,Zic82}. This is done as follows: The boundary
  condition for the fluid is given by a unique constant (but unknown)
  velocity on the surface. To assure this boundary condition, one may
  introduce an artificial ``reaction force density''
  $\boldsymbol{f}_{reac}$ located on the surface (with units: force
  per area). This force density needs to be determined
  self--consistently such that the superposition of the flow fields
  generated by the external forces and these reaction forces satisfies
  the boundary condition. This is in spirit quite analogous to the
  electrostatic problem of a metallic surface, where the problem of
  finding a constant electrostatic potential at the surface is solved
  by determining an appropriate induced charge density. It is clear
  that this reaction force density cannot exert a net force onto the
  system, and hence we know
  \begin{equation}
    \label{eq:no_total_reaction_force}
    \int d \Omega \boldsymbol{f}_{reac} (\boldsymbol{r}) = 0 ;
  \end{equation}
  here $\Omega$ denotes the surface.

  We thus can write the total flow field $\boldsymbol{v}$ as a
  superposition of $\boldsymbol{v}_1$, the contribution from the
  external force density, and $\boldsymbol{v}_2$ coming from the
  reaction force density:
  \begin{eqnarray}
      \boldsymbol{v} (\boldsymbol{r}) & = & 
      \boldsymbol{v}_1 (\boldsymbol{r}) +
      \boldsymbol{v}_2 (\boldsymbol{r}) \\
      \boldsymbol{v}_1 (\boldsymbol{r}) & = & 
      \int_V d^3 r' \, \tensor{\boldsymbol{T}} ( \boldsymbol{r}
      - \boldsymbol{r}')
      \, \boldsymbol{f}_{ext} (\boldsymbol{r}') \,,\\
      \boldsymbol{v}_2 (\boldsymbol{r}) & = & 
      \int d \Omega' \, \tensor{\boldsymbol{T}} ( \boldsymbol{r}
      - \boldsymbol{r}')
      \, \boldsymbol{f}_{reac} (\boldsymbol{r}') \, ,
  \end{eqnarray}
  where $\tensor{\boldsymbol{T}}$ is the Green's function of the
  Stokes equation.

  For an infinite fluid this Green's function is well known and given
  by the Oseen tensor (see e.~g. \cite{LL6}). In Fourier space it is
  given by
  \begin{equation} \label{eq:Oseen_Fourier}
      {\hat{\tensor{\boldsymbol{T}}}}
      = \frac{1}{\eta \boldsymbol{k}^2}
      \left( \tensor{\boldsymbol{I}}
      - \frac{\boldsymbol{k} \otimes
      \boldsymbol{k}}{\boldsymbol{k}^2} \right) \, ,
  \end{equation}
  where $\boldsymbol{I}$ denotes the unit tensor. In a finite box, the
  same form still applies; however, one needs to take into account
  that only wave vectors $\boldsymbol{k}$ occur that are compatible
  with the box periodicity. Furthermore, $\boldsymbol{k} = 0$ must be
  excluded, since the problem is solved in the system's
  center--of--mass reference frame. This yields for the real--space
  counterpart
  \begin{equation}\label{eq:Oseen_Real}
      \tensor{\boldsymbol{T}} (\boldsymbol{r})
      = \frac{1}{V \eta} \sum_{\boldsymbol{k} \neq 0}
	\frac{\exp[-i\boldsymbol{k} \cdot\boldsymbol{r}]}
	{\boldsymbol{k}^2}
	\left( \tensor{\boldsymbol{I}}
	- \frac{\boldsymbol{k} \otimes
	\boldsymbol{k}}{\boldsymbol{k}^2} \right)\,;
  \end{equation}
  here $V$ denotes the total volume of the box. It should be noted
  that we can thus consider $\boldsymbol{v}_1$ as known, while we do
  not know $\boldsymbol{v}_2$ and $\boldsymbol{f}_{reac}$.

  Now, picking two points $\boldsymbol{r}_\Omega$ and $\boldsymbol{r}_{ref}$
  that are both located on the surface, we know that their velocity
  must be identical:
  \begin{eqnarray}
    \boldsymbol{v}_1 (\boldsymbol{r}_\Omega) +
    \boldsymbol{v}_2 (\boldsymbol{r}_\Omega)
    & = &
    \boldsymbol{v}_1 (\boldsymbol{r}_{ref}) +
    \boldsymbol{v}_2 (\boldsymbol{r}_{ref}) ,
    \\
    \boldsymbol{v}_2 (\boldsymbol{r}_\Omega) -
    \boldsymbol{v}_2 (\boldsymbol{r}_{ref})
    & = &
    \boldsymbol{v}_1 (\boldsymbol{r}_{ref}) -
    \boldsymbol{v}_1 (\boldsymbol{r}_\Omega) ,
    \end{eqnarray}
    \begin{eqnarray}
    \nonumber
    &&
    \int d \Omega' 
      \left( \tensor{\boldsymbol{T}} ( \boldsymbol{r}_\Omega
      - \boldsymbol{r}_{\Omega'})
      - \tensor{\boldsymbol{T}} ( \boldsymbol{r}_{ref}
      - \boldsymbol{r}_{\Omega'}) \right)
      \boldsymbol{f}_{reac} (\boldsymbol{r}_{\Omega'}) \\
    & = &
    \boldsymbol{v}_1 (\boldsymbol{r}_{ref}) -
    \boldsymbol{v}_1 (\boldsymbol{r}_\Omega) .
    \label{eq:determine_f_reac}
  \end{eqnarray}
  In a discretised version the $\Omega'$ integral is replaced by a
  sum. If we now view $\boldsymbol{r}_{ref}$ as an arbitrary but fixed
  reference point, while we vary $\boldsymbol{r}_\Omega$, we may view
  Eq. \ref{eq:determine_f_reac} as a system of linear equations to
  determine $\boldsymbol{f}_{reac}$. If the number of surface points
  is $M$, then the number of equations is $3M$, while the number of
  unknowns is $3M$ as well. However, three of these equations are
  redundant, since at the point $\boldsymbol{r}_\Omega =
  \boldsymbol{r}_{ref}$ only the trivial information $0 = 0$ is
  obtained. Instead, we need to use
  Eq. \ref{eq:no_total_reaction_force} as last set of equations to
  obtain a unique solution. In practice, the set of equations was
  solved numerically using the standard BiCGStab procedure
  \cite{IML96,Mei99}.

  For discretisation, we again use a simple--cubic lattice with
  spacing $a$. For consistency reasons, we need to use the discrete
  version of the Oseen tensor, analogously to the lattice Green's
  function of the Poisson equation (see Eqs.
  \ref{eq:DiscretePoisson_1} and \ref{eq:DiscretePoisson_2} and
  Ref. \cite{NumRec}). The discretised Oseen tensor is derived by
  applying a midstep finite--difference scheme in real space, and
  doing the corresponding discrete Fourier transform with integer
  indexes $k,l,m$:
  \begin{equation}
      \hat{\tensor{\boldsymbol{T}}} (k,l,m)
      \,=\, \frac{1}{\eta \xi_{k,l,m}^2}
	\left(\tensor{\boldsymbol{I}}
	- \frac{\boldsymbol{q}_{k,l,m}
	\otimes \boldsymbol{q}_{k,l,m}}
	{\boldsymbol{q}_{k,l,m}^2} \right)
  \end{equation}
  with
  \begin{eqnarray}
      \boldsymbol{q}_{k,l,m} & = &
      \frac{1}{a} \left\{
      \sin\left(2\pi \frac{k}{N_x}\right) ,
      \sin\left(2\pi \frac{l}{N_y}\right) ,
      \sin\left(2\pi \frac{m}{N_z}\right)
      \right\} \, , \\
      \xi_{k,l,m}^2  & = & \frac{2}{a^2} \left\{3 -
	\cos\left(2\pi \frac{k}{N_x}\right)
      - \cos\left(2\pi \frac{l}{N_y}\right)
      - \cos\left(2\pi \frac{m}{N_z}\right) \right\} \, .
  \end{eqnarray}

  \subsection{Convection-diffusion equation}
  \label{ssec:CDE}

  Equation \ref{eq: first order EKE 2 a} is the stationary limit of a
  convection-diffusion equation of the general form
  \begin{equation} \label{eq: orig_convdiff}
    \left( \frac{\partial}{\partial t} 
    + \nabla \cdot {\boldsymbol{w}} (\boldsymbol{r}) \right)
    c(\boldsymbol{r}, t) =
    D \nabla^2 c(\boldsymbol{r}, t)  + S(\boldsymbol{r}, t) \,.
  \end{equation}
  Here $c(\boldsymbol{r},t)$ denotes the ionic concentration field of
  first order at the spatial position $\boldsymbol{r}$ and time
  $t$. The convective term is not the fluid velocity, but rather the
  coupling term of the first order ion concentration with the zeroth
  order electrostatic potential, $\boldsymbol{w}(\boldsymbol{r}) = - D
  z \nabla \psi^{(0)} (\boldsymbol{r})$. The source term
  $S(\boldsymbol{r},t)$ contains all other terms independent of the
  first--order ion concentration field. Equation \ref{eq: first order
    EKE 2 a} is a conservation law, and therefore
  \begin{eqnarray}
      \int d^3 r \, c(\boldsymbol{r},t) & = & \mbox{\rm const.} \,, \\
      \int d^3 r \, S(\boldsymbol{r},t) & = & 0 \, .
  \end{eqnarray}
  The latter equation states that no ionic particles are produced or
  annihilated during the process. The discrete counterpart of such an
  equation is a Master equation, the coefficients of which need to be
  adjusted such that its continuum limit recovers Eq. \ref{eq:
    orig_convdiff}. A detailed and systematic derivation of such an
  algorithm shall be published separately \cite{RS2011}. In the
  present work, we use the simplest version, which is a
  nearest--neighbour model on the simple--cubic lattice, which has
  lowest--order accuracy, and present the algorithm without proof. The
  concentration fields are initialised as zero at every lattice site
  and then propagated using a Master equation of the form
  \begin{equation}
      \nonumber
      c(\boldsymbol{r},t) = 
      \sum_i A_i ( \boldsymbol{r} - a \boldsymbol{\Delta}_i) \,
      c ( \boldsymbol{r} - a \boldsymbol{\Delta}_i, t - h)
      + h S(\boldsymbol{r},t - h) \,,
  \end{equation}
  where $h$ is the discretisation time step, $\boldsymbol{r}$ denotes
  the lattice sites and $a$ is the lattice
  spacing. $\boldsymbol{\Delta}_i$ denotes a dimensionless lattice
  vector connecting $\boldsymbol{r}$ with one of its neighbours, such
  that $\boldsymbol{r} + a \boldsymbol{\Delta}_i$ is again a lattice
  site. For our simple three--dimensional nearest--neighbour model, $i
  = 1, \dots, 6$, and the transfer coefficients are given by
  \begin{equation}
      A_i (\boldsymbol{r} ) \, = \, \frac{1}{6} \left(
      1 + \frac{a}{2 D} 
      \boldsymbol{w} (\boldsymbol{r}) \cdot \boldsymbol{\Delta}_i
      \right) \, ,
  \end{equation}
  while the diffusion constant is
  \begin{equation}
      D = \frac{1}{6} \frac{a^2}{h} \, .
  \end{equation}
  This means that the continuum limit is obtained by Taylor expansion up
  to second order with respect to only one variable $a$, since the
  time step is not a variable that can be picked independently from
  $a$, but rather is given by $h = a^2 / (6D)$.

  \section{Numerical results}
  \label{sec:results}

  \subsection{Parameters}

  From now on, we will return to the notation of
  Sec. \ref{ssec:dimless} and, for reasons of clarity, distinguish
  between dimensional and reduced quantities.

  The electrophoretic mobility $\mu_{red}$ is a dimensionless
  quantity, and hence it can only depend on dimensionless parameters
  as well. In Ref. \cite{Due08} we discussed, within the framework of
  the present Mean--Field treatment, a finite system with added salt,
  where all ion types have the same properties, i.~e. all ions are
  monovalent and have all the same friction coefficient. We then found
  as one possible set of dimensionless parameters: (i) the reduced
  charge
  \begin{equation} \label{hatZdefinition}
    \hat Z \, = \, Z \frac{l_B}{R} \, = \,
    \frac{\tilde Z}{4 \pi \tilde R} \, ,
  \end{equation}
  (ii) the rescaled colloid radius $\tilde R = \kappa R$, (iii) the
  rescaled diffusion constant of the ions $\tilde D$ (as defined in
  Tab. \ref{Tab:Reduced Parameters}) and (iv) a dimensionless quantity
  $f_0$ that specifies the fraction of counterions (species zero)
  relative to the salt ions. In general $f_i$ is the fraction of the
  ionic species $i$ relative to all ions in the system,
  \begin{equation} \label{eq: fi definition}
      f_i \,=\, \frac{N_i}{\sum_j z_j^2 N_j} \, ,
  \end{equation}
  and it is easily shown that $f_i$ is nothing but the
  volume--averaged concentration of ionic species $i$, in the reduced
  unit system of Tab. \ref{Tab:Reduced Parameters}. For a system with
  only one salt species that has only monovalent ions, we know $f_1 =
  f_2$ (these fractions refer to the salt ions), and the sum rule
  $\sum_i z_i^2 f_i = 1$ implies that only one non--trivial parameter
  $f_0$ is left.

  As already mentioned in Sec. \ref{ssec:dimless}, a definition of
  $\kappa$ that is fully consistent with the finite--volume version
  of linearized Poisson--Boltzmann theory requires that the volume
  $V$ is defined as the volume available to the ions. For a box of
  dimension $L \times L \times L$ and $N$ colloidal spheres of
  radius $R$ this means
  \begin{equation}
    V = L^3 - \frac{4 \pi}{3} N R^3 \, .
  \end{equation}
  In our notation, $N$ is the number of colloids and $Z$ is their
  charge, while $N_0$ and $z_0$ are the number and valence of
  counterions, respectively, such that charge neutrality implies 
  \begin{equation} \label{eq:NZn0z0}
    N Z = - N_0 z_0 .
  \end{equation}
  The colloid volume fraction is thus given by
  \begin{equation}
    \Phi \, = \, 
    \frac{\frac{4 \pi}{3} N R^3}{V + \frac{4 \pi}{3} N R^3}
  \end{equation}
  or
  \begin{equation} \label{eq:volfrac_renorm}
    \frac{4 \pi}{3} \frac{N R^3}{V} = \frac{\Phi}{1 - \Phi} .
  \end{equation}
  Therefore the correct relation between volume fraction $\Phi$ and
  $f_0$ differs from the expression in Ref. \cite{Due08} by a small
  correction term. Inserting Eqs. \ref{eq:Debye_screening},
  \ref{hatZdefinition}, \ref{eq: fi definition} and \ref{eq:NZn0z0},
  one finds after a few lines of algebra
  \begin{equation} \label{eq:Phi_vs_f0}
    \frac{\Phi}{1 - \Phi} \, = \, 
    - \frac{z_0 f_0}{3 \hat Z} (\kappa R)^2
    \, = \,
    - \frac{z_0 f_0}{3 \hat Z} \tilde{R}^2 \, .
  \end{equation}

  Furthermore, a dimensionless resolution $d$ is defined such that for
  given $d$ a sphere is always discretised by the same number of
  lattice sites:
  \begin{equation}
     d = \frac{\tilde{a}}{\tilde R} \, ,
  \end{equation}
  where $\tilde{a}$ denotes the lattice spacing in reduced units and
  $\tilde R$ is the radius of the particle, also in reduced units.

  \subsection{Comparison with previous results}

  \begin{figure}
    \begin{center}
      \includegraphics[clip,width=0.7\textwidth]
      {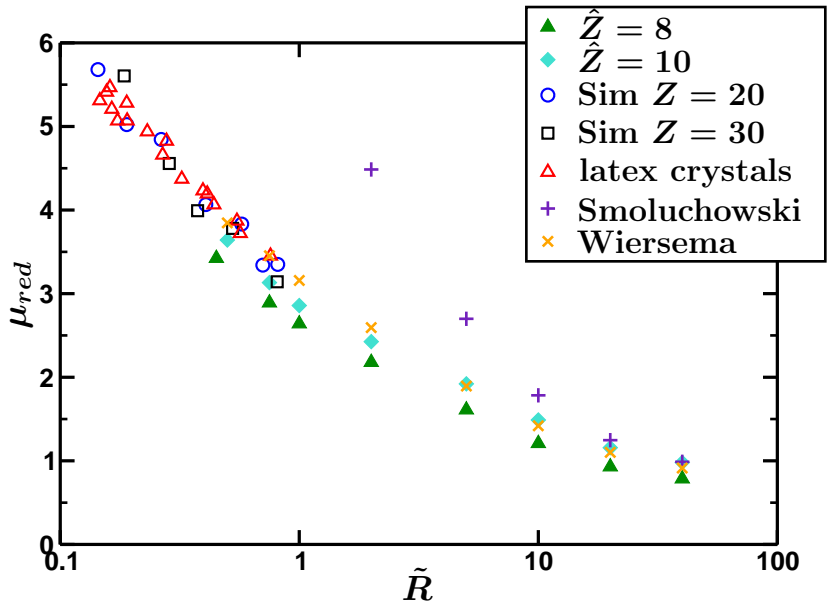}
    \end{center}
      \caption{Reduced mobility as a function of $\tilde R$.
	       The open symbols are results from Ref. \cite{Lobaskin2007}.}
    \label{fig:compare_Lobaskin}
  \end{figure}

  Figure \ref{fig:compare_Lobaskin} studies the reduced mobility for a
  system consisting of one colloidal sphere in a box, where the volume
  fraction $\Phi \simeq 7.07 \cdot 10^{-3}$ as well as the reduced
  charge $\hat{Z} = 8,10$ are kept fixed. The counterions are
  monovalent. $\tilde R$ is varied by changing $f_0$
  (cf. Eq. \ref{eq:Phi_vs_f0}), i.~e., by varying the salt content
  (one species, all salt ions monovalent). The calculations were done
  with a reduced diffusion constant $\tilde D_i = 1.5$ for all ionic
  species, and the resolution was kept fixed at the value $d =
  0.07$. One clearly sees that the mobility systematically decreases
  with $\tilde R$, which is easily explained by the corresponding
  increase of electrostatic screening. Note that the present
  representation is given for constant $\hat Z$, which differs from
  the classical calculations \cite{Ove43,Boo50,Wie66,Ohs83,OBr78} that
  keep the zeta potential fixed, i.~e. the electrostatic potential at
  the colloidal surface.

  Furthermore, Fig. \ref{fig:compare_Lobaskin} shows also data from
  Ref. \cite{Lobaskin2007} (open symbols). The circles and squares are
  simulation results obtained from the Molecular Dynamics / lattice
  Boltzmann (MD/LB) raspberry model \cite{Lob2004}. In the simulations
  a single colloid of charge $Z = 20$ ($Z = 30$) and radius $R_C = 3$
  ($R_C = 5$) is surrounded by counterions. Both systems were studied
  with a Bjerrum length of $l_B = 1.3$, resulting in a reduced charge
  of $\hat Z = 8.5$, comparable to our value. The triangles are
  experimental results for latex crystals in a bcc structure with a
  particle size of $R = 34nm$ in a deionised aqueous suspension. The
  effective charge is quoted as $Z \simeq 450$ determined from
  conductivity measurements \cite{Wet02}, resulting in a reduced
  charge of order $\hat Z \simeq 10$. In all cases of
  Ref. \cite{Lobaskin2007} the colloidal size and charge were fixed
  and salt ions were absent (except for the self--dissociation of
  water). The screening parameter, or $\tilde R$, was hence changed by
  varying $\Phi$, while $f_0$ was kept constant (cf. again
  Eq. \ref{eq:Phi_vs_f0}). This is qualitatively different from our
  numerical calculations, where rather $\Phi$ is kept constant and
  $f_0$ is varied. Nevertheless, the results seem to agree quite
  nicely, and within the accuracy of the data it seems that it does
  not matter whether the screening is salt--dominated or
  counterion--dominated. In the following subsection, we will put this
  question under more detailed and more accurate scrutiny.

  Besides the question of the screening mechanism, there are also
  additional differences between our calculation and
  Ref. \cite{Lobaskin2007}, in view of which the observed small
  discrepancies are hardly surprising: The MD/LB simulations use a
  slightly different reduced charge, and also the reduced diffusion
  constant of the ions is probably somewhat different. The influence
  of the diffusion constants on $\mu$ has so far not been thoroughly
  studied; in the next subsection it will be shown that $\mu$
  increases with the $D_i$, as one might expect. For the experiments,
  the situation is even less clear, since there is a considerable
  amount of ambiguity in the mapping of the effective charge of the
  real physical system to the bare charge in the Mean-Field
  calculations.

  Finally, Fig. \ref{fig:compare_Lobaskin} presents also a comparison
  with two theoretical results. Firstly, the Smoluchowski limit
  \cite{Smo03} of the reduced mobility is given by
  \begin{equation}
     \mu_{red}^{Smo} = \frac{3}{2} \zeta_{red} \, ,
  \end{equation}
  where $\zeta_{red}$ is the reduced (dimensionless,
  cf. Tab. \ref{Tab:Reduced Parameters}) zeta potential, i.~e.  the
  electrostatic potential at the colloid surface. This can be easily
  calculated within our approach; it is just a result of our
  Poisson--Boltzmann solver, where we take for $\zeta$ the potential
  difference between colloid surface and box boundary. Secondly,
  within an approximate numerical theory for a single colloid in an
  infinite salt solution, Wiersema et al. \cite{Wie66} have tabulated
  values for $\mu_{red}$ as a function of $\zeta_{red}$ and $\tilde
  R$, while the influence of $\tilde D_i$ is stated there to be fairly
  small. We can therefore use our values for $\tilde R$ and
  $\zeta_{red}$ to also compare with that theory, using linear
  interpolation. While the Smoluchowski limit is reached for values
  $\tilde R \simeq 40$, the data obtained from the work of Wiersema et
  al. describe our numerical results quite well over the full range.

  \begin{figure}
    \begin{center}
      \includegraphics[clip,width=0.7\textwidth]
      {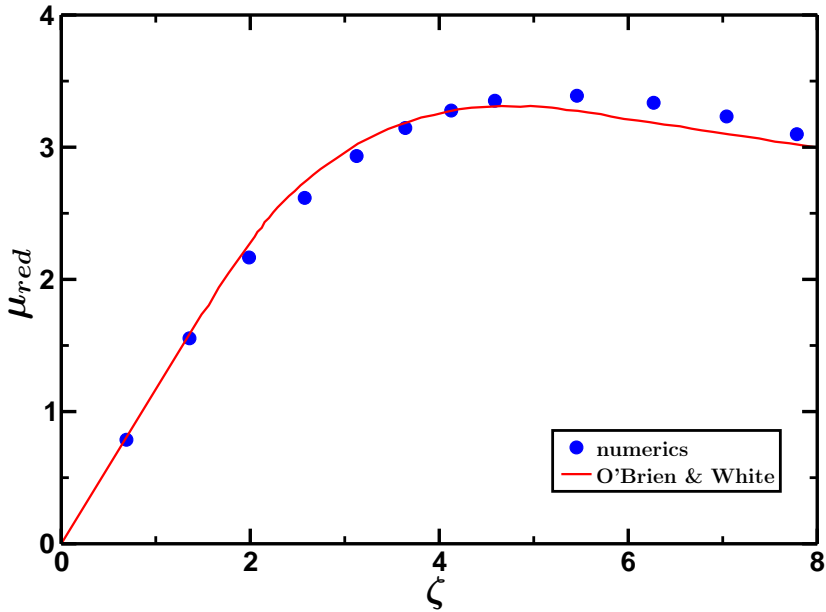}
    \end{center}
    \caption{Reduced mobility as a function of the dimensionless
      (reduced) zeta potential (determined from the solution of the
      Poisson--Boltzmann equation) for $\tilde R = 8$, $d = 0.07$,
      $\tilde D = 1$ and $f_0 \simeq 0.01$. The solid line is the
      result taken from Ref. \cite{OBr78}.}
    \label{fig:ObrienWhite}
  \end{figure}

  Since our method is the extension of O'Brien and White's work
  \cite{OBr78} to systems with finite volume fraction and a finite
  amount of counterions, it should produce identical results in the
  limit of strongly salt--dominated screening ($f_0 \to 0$), where the
  ionic cloud of the colloid does not overlap with those of its
  periodic images. A quantitative comparison is however hampered by
  the fact that Ref. \cite{OBr78} does not tabulate its results, but
  only provides plots, and, more importantly, that the values of the
  ionic diffusion constants are not quoted. It seems however that
  $\tilde D = 1$ for all ionic species is reasonable; at least we do
  find quite good agreement between our calculations and
  Ref. \cite{OBr78} for this value, as demonstrated in Fig.
  \ref{fig:ObrienWhite}.

  \subsection{Salt vs. counterion screening}
  \label{ssec:salt}

  The parameter $f_0$ can be used to quantify the screening
  mechanism. Values close to unity (for a monovalent system) indicate
  a system where the amount of counterions in the system is dominant,
  while a value close to zero means that the salt ions dominate the
  screening mechanism. In order to focus on the screening
  \emph{mechanism} (salt screening vs. counterion screening), one
  should keep $\kappa$ (or $\kappa R$) strictly constant, while
  varying \emph{only} $f_0$. Within our numerical approach, this is
  easily possible: The computer experiment consists of adding more and
  more salt (in terms of concentration), which enhances the screening,
  while at the same time the box size is increased, such that the
  counterions are more and more diluted and their contribution to the
  screening is reduced. This is done in such a way that the total
  amount of screening remains constant (see Eq. \ref{eq:Phi_vs_f0}).
  Of course, the reduced charge and the reduced diffusion coefficients
  are kept constant as well.

  A common assumption is that only the screening length, but not the
  screening mechanism should contribute to the value of the
  electrophoretic mobility. Although Fig. \ref{fig:compare_Lobaskin}
  and previous studies \cite{Lobaskin2007} show that within the given
  accuracy the effect of $f_0$ on the mobility is at least weak, there
  is no fundamental reason why the mobility should be strictly
  independent of $f_0$. 

\begin{figure}
    \begin{center}
      \includegraphics[clip,width=0.7\textwidth]{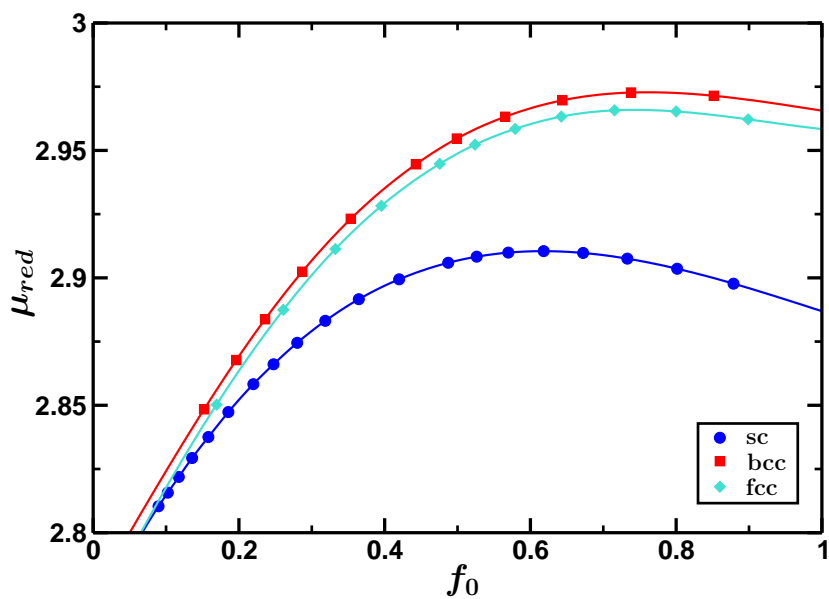}
    \end{center}
    \caption{Reduced mobility as a function of $f_0$ for constant
      $\tilde R = 1$ and three types of lattice structure as indicated
      by the legend. Every sphere carries a reduced charge of $\hat Z
      = 10$ and the computational resolution is $d = 0.08$.}
    \label{fig:Mobility_f0}
  \end{figure}

  In order to test this quantitatively, we have studied (i) a single
  colloidal sphere in a box, corresponding to a simple--cubic (sc)
  crystal, (ii) two spheres in the box, such that the resulting
  crystal is body--centered cubic (bcc), and (iii) four spheres
  arranged in such a way to construct a face--centered cubic (fcc)
  crystal. The fixed parameters were $\tilde R = 1$, $\hat Z = 10$,
  $\tilde D = 1.5$, $d=0.08$, while $f_0$ was varied. Since we work at
  constant resolution, the resulting curves ($\mu_{red}$ vs. $f_0$) in
  Fig. \ref{fig:Mobility_f0} are smooth. Indeed, the plot nicely shows
  that the reduced mobility does depend on the screening mechanism in
  a non--trivial fashion, for all three types of lattice
  structures. However, the effect is only of the order of $5$ to $10
  \%$, and hence was not observable previously.

  \begin{figure}
    \begin{center}
      \includegraphics[clip,width=0.7\textwidth]
      {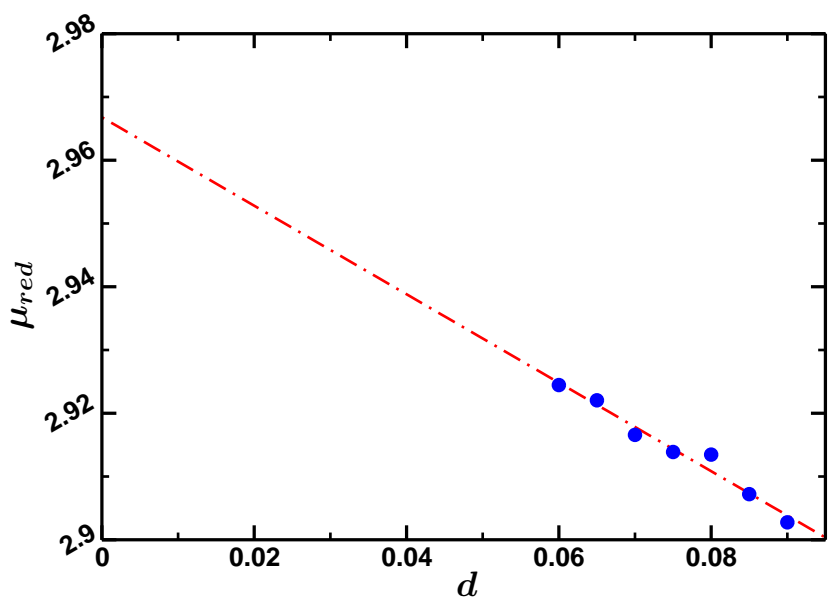}
    \end{center}
    \caption{Reduced electrophoretic mobility as a function of the
      sphere resolution $d$ for a constant value of $f_0 = 0.5$. The
      red line shows a linear fit to the numerical data.}
    \label{fig:Mobility_resolution}
  \end{figure}

  \begin{figure}
    \begin{center}
      \includegraphics[clip,width=0.7\textwidth]
      {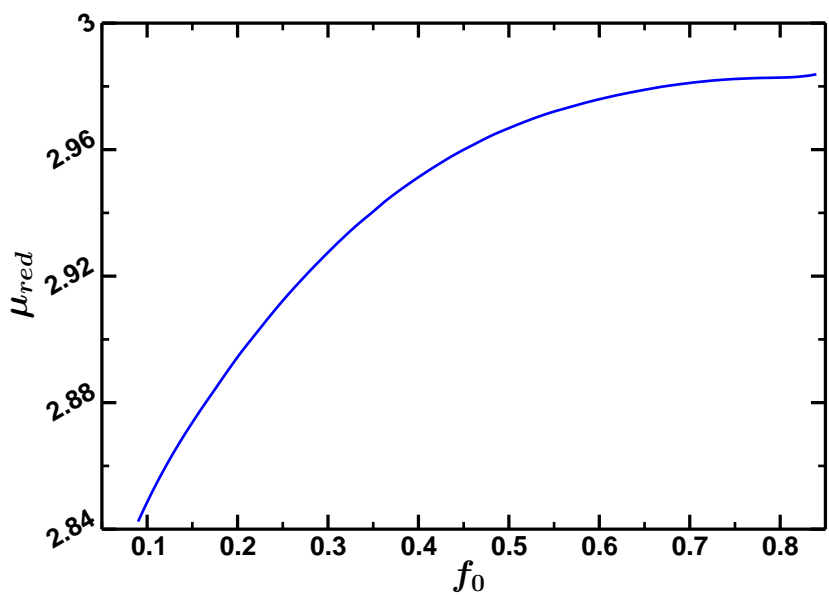}
    \end{center}
    \caption{Reduced electrophoretic mobility as a function of $f_0$
      in the continuum limit $d \to 0$.}
    \label{fig:Mobility_continuumslimes}
  \end{figure}

  \begin{figure}
    \begin{center}
      \includegraphics[clip,width=0.7\textwidth]
      {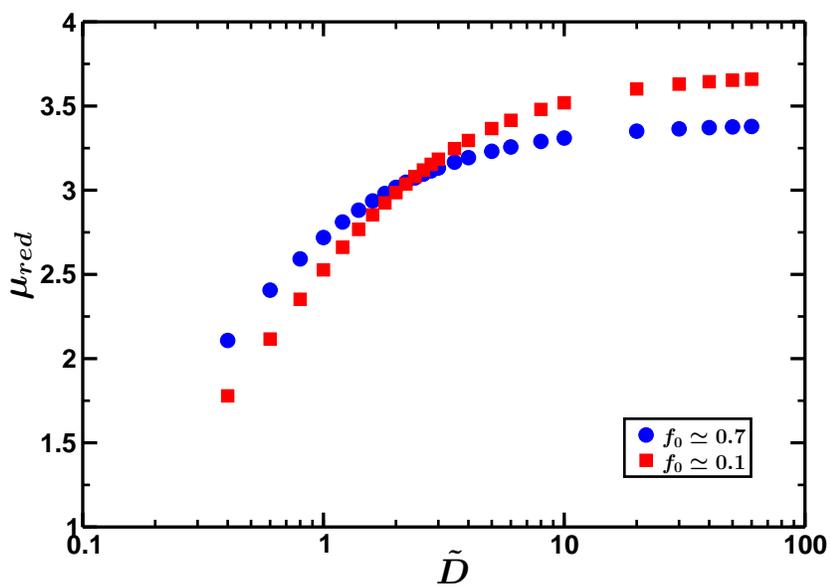}
    \end{center}
    \caption{Reduced electrophoretic mobility as a function of the
      diffusion constant. Both curves are generated for a system with
      $\tilde R = 1$, $\hat Z = 10$, $d = 0.07$ and various values of
      $f_0$ as indicated by the legend.}
    \label{fig:Mobility_Diffusion}
  \end{figure}

  \begin{figure}
    \begin{center}
      \includegraphics[clip,width=0.7\textwidth]
      {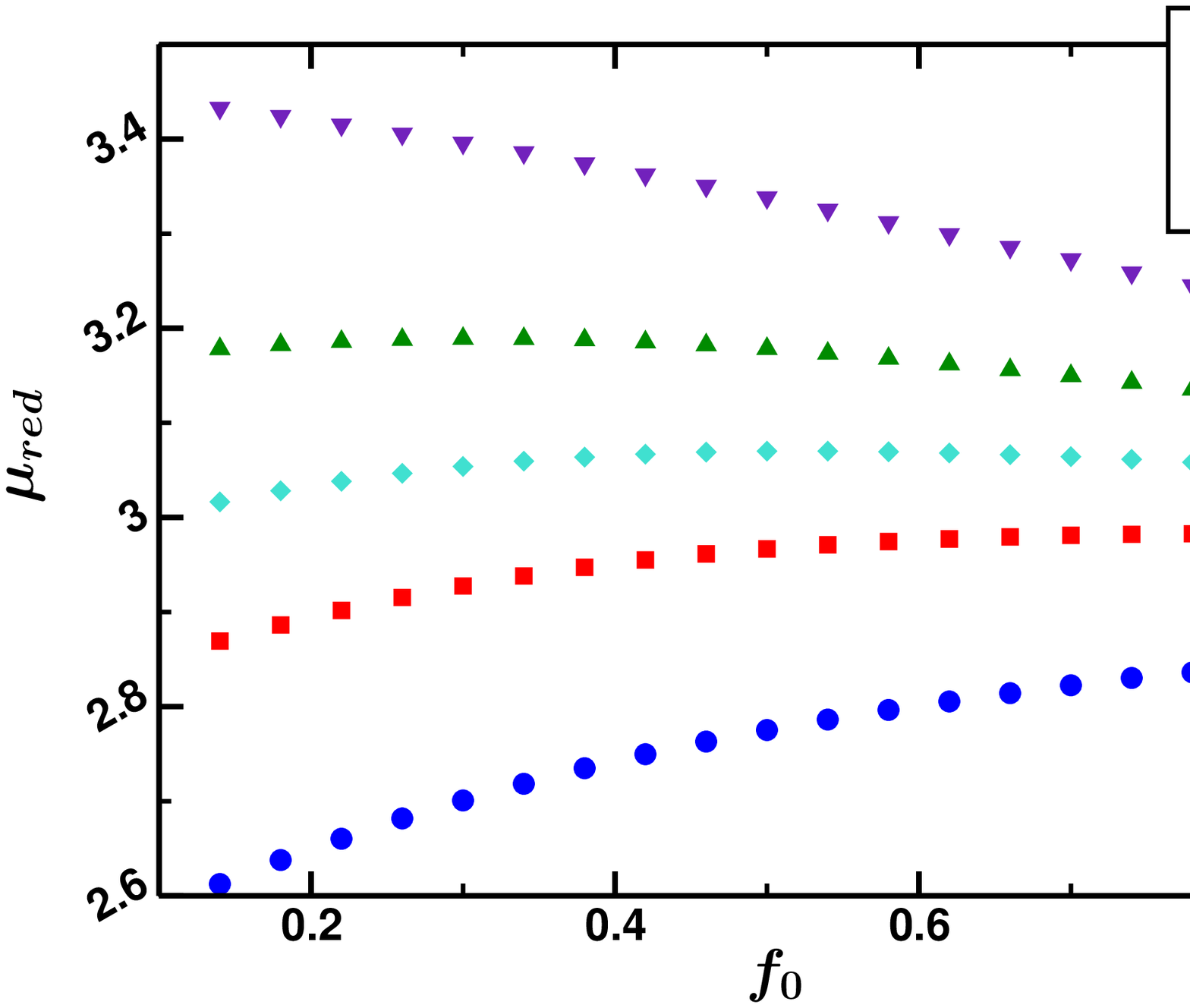}
    \end{center}
    \caption{Reduced mobility as a function of $f_0$ for one single
      sphere in a periodic box (sc lattice) and the same parameters as
      in Fig. \ref{fig:Mobility_f0} for different diffusion constants
      as indicated by the legend. This plot shows $d\to 0$
      extrapolated values.}
    \label{fig:Mobility_f0_diffusion}
  \end{figure}

  In order to assess the effect of the lattice resolution on this
  result, we analysed the sc case in some more detail by varying $d$
  as well. Due to computational limitations, this was however confined
  to a fairly narrow interval $0.06 \le d \le 0.09$. Nevertheless, a
  reasonably reliable linear extrapolation to the continuum limit $d
  \to 0$ seems possible, see Fig. \ref{fig:Mobility_resolution}, where
  this is shown for the data point $f_0 = 0.5$. The result of this
  extrapolation, namely the reduced mobility as a function of $f_0$ in
  the limit $d \to 0$, is presented in Fig.
  \ref{fig:Mobility_continuumslimes}, where cubic splines were used
  for interpolation. Even though the data may not be fully reliable
  due to the smallness of the $d$ interval, they nevertheless indicate
  fairly convincingly that the effect is more than just a
  discretisation artifact.

  Furthermore, it turns out that this non--trivial behaviour is even
  more interesting when studying the effect of the diffusion
  constants. The limiting behaviour for $D_i \to 0$ and for $D_i \to
  \infty$ can be easily understood. For $D_i \to 0$, only the
  convective part of the convection--diffusion equation remains:
  \begin{equation}
     \boldsymbol{v}^{(1)} 
     \cdot \nabla c_i^{(0)} \, = \, 0\, .
  \end{equation}
  At least for a single sphere in infinite space it is easily shown
  that this enforces the trivial solution $\boldsymbol{v}^{(1)} = 0$
  (and hence $\mu = 0$): In a spherical coordinate system with origin
  at the center of the sphere, and polar angle relative to the driving
  field, it is clear that the radial component of
  $\boldsymbol{v}^{(1)}$ must vanish, due to the above condition,
  while the azimuthal component must be zero as well, due to
  symmetry. Therefore only the polar component remains. However,
  imposing the condition $\nabla \cdot \boldsymbol{v}^{(1)} = 0$ for
  that component yields a solution that is either singular (and hence
  forbidden) or zero. We hence find $\mu \to 0$ for $D_i \to 0$, and
  it is highly plausible that this holds in the general case as
  well. Conversely, the limit $D_i \to \infty$ means that the
  convective term can be ignored, and the problem becomes independent
  of $D_i$. Hence the electrophoretic mobility saturates at some
  limiting value.

  These predictions are nicely confirmed in our calculations, see
  Fig. \ref{fig:Mobility_Diffusion}. The data shown there were
  calculated for $\tilde R = 1$, $\hat Z = 10$ and $d = 0.07$, but for
  different screening mechanisms. Since the curves intersect, one sees
  that switching from salt--dominated to counterion--dominated
  screening may either enhance the mobility (this happens for small
  diffusion coefficients) or reduce it (this is the behaviour at large
  $D$). In more detail, this behaviour is analysed in
  Fig. \ref{fig:Mobility_f0_diffusion}.

  \subsection{Weakly charged colloids}
  \label{ssec:weak_charge}

  A very interesting phenomenon can be observed in the case of weakly
  charged colloids. Consider an uncharged spherical obstacle in a
  solution of negatively and positively charged ions. Applying a
  constant external electric field, electro--osmotic flow is generated
  by electric forces acting on the salt ions; positive charges move
  with the field direction, negative charged ions against it. Since
  the ions can not penetrate the solid sphere, the ion fluxes will be
  deflected by the particle. Thus, negative salt ions accumulate at
  one side of the sphere, while positive ions are depleted in that
  region. Since no electric forces act on the uncharged particle, the
  accumulation of positive ions at one side and negative ions at the
  opposite side must be symmetric. This accumulation effect leads to a
  polarisation of the system. Note that the induced dipole moment
  points in the ``wrong'' direction, i.~e. anti--parallel to the
  driving field. This is interesting, because it is in contrast to the
  usual observation that for a charged sphere the induced dipole
  moment points in the ``right'' direction, i.~e. parallel to the
  external field \cite{Lobaskin2004}.

\begin{figure}
    \begin{center}
      \includegraphics[clip,width=0.7\textwidth]
      {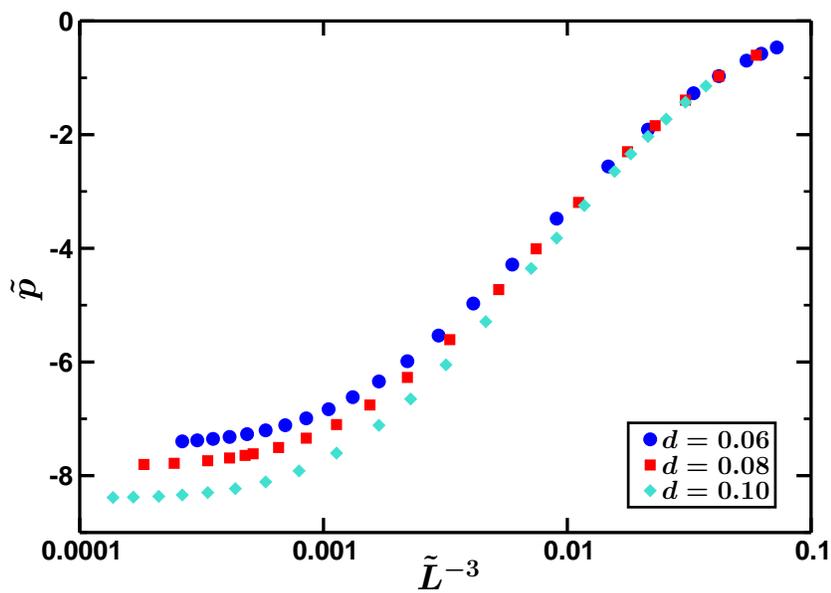}
    \end{center}
    \caption{First order dipole moment as a function of the inverse
      box size for various resolutions as indicated by the legend.}
    \label{fig:Dipole_boxsize}
  \end{figure}

\begin{figure}
    \begin{center}
      \includegraphics[clip,width=0.7\textwidth]
      {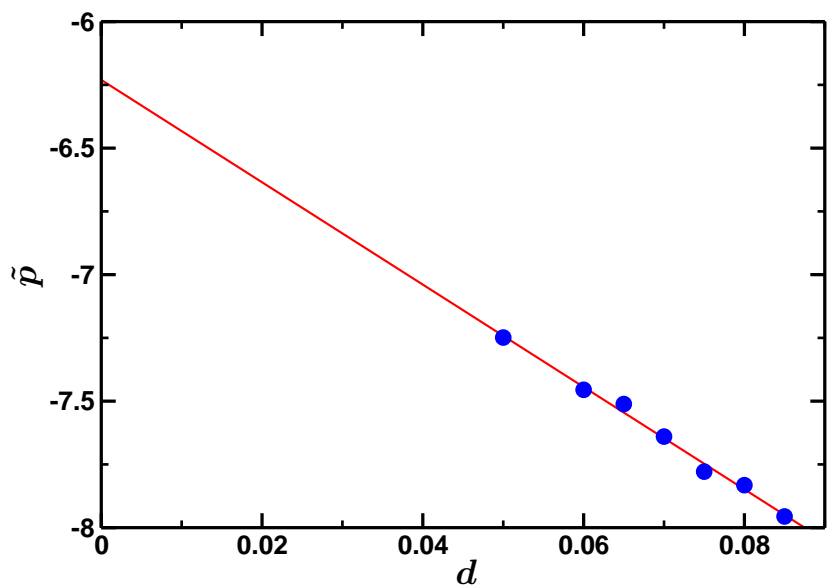}
    \end{center}
    \caption{Dipole moment as a function of the resolution $d$,
     where the data points are the results of extrapolations
     to the $L \to \infty$ limit.}
    \label{fig:Dipole_resolution}
  \end{figure}

  For an uncharged sphere in an infinite salt solution, the problem is
  dramatically simplified and amenable to an exact analytical
  solution; this was recently presented by Dhont and Kang
  \cite{Dhont2010}. In reduced units their result for the dipole
  moment is
  \begin{equation}\label{eq:dipole_analytic}
      \tilde{\boldsymbol{p}} = - 2 \pi {\tilde R}^3 
      \tilde{\boldsymbol{E}}_{ext} \, .
  \end{equation}
  Our numerical calculations nicely reproduce this prediction for a
  system with $\hat Z = 0$, $\tilde R = 1$, $\tilde E = 1$. However,
  again we need to reach the limit of an infinite system, meaning
  $\Phi \to 0$, as well as the continuum limit $d \to 0$. This double
  extrapolation is presented in Figs.  \ref{fig:Dipole_boxsize} and
  \ref{fig:Dipole_resolution}, and our final numerical result is
  \begin{equation}
     \tilde p \simeq - 6.23 \, ,
  \end{equation}
  which deviates less than $1 \%$ from the expected value $- 2 \pi$.

  \begin{figure}
    \begin{center}
      \includegraphics[clip,width=0.7\textwidth]
      {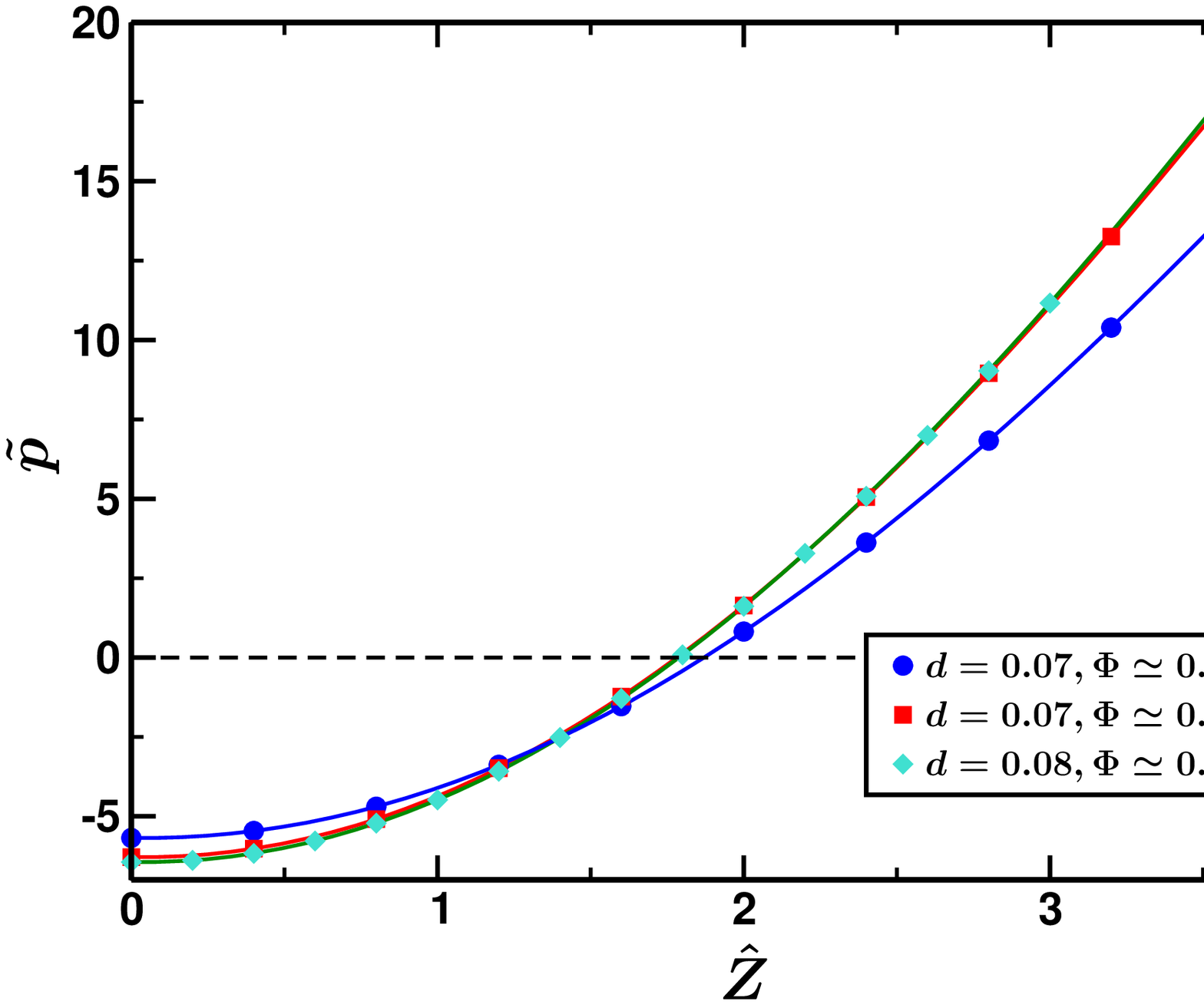}
    \end{center}
    \caption{Dipole moment as a function of the reduced particle charge,
     for $\tilde R = 1$, $\tilde E = 1$, and various volume fractions
     and resolutions as indicated.}
    \label{fig: dipole vs charge}
  \end{figure}

  In a second step, the reduced charge $\hat Z$ of the colloidal
  sphere was increased, while we kept $\tilde R = 1$, $\tilde E = 1$
  and used $\tilde D = 1.5$. The results for the dipole moment are
  presented in Fig. \ref{fig: dipole vs charge}, and they will be
  discussed below.

  We would like to comment that we believe that at this point the
  advantages of our perturbative approach come to full effect. In a
  non--perturbative calculation, the induced dipole moment would be a
  very weak signal on top of the leading--order charge cloud, and this
  weak contribution may be fairly difficult to be disentangled from
  artifacts in the leading order (discretisation errors and roundoff
  errors which result in an artificial nonzero dipole moment).
  Conversely, in our calculation the leading--order and the
  first--order contribution are cleanly separated. We hence believe
  our method to be more accurate and stable than non--perturbative
  approaches.

  \begin{figure}
    \begin{center}
      \begin{tabular}{c c c}
       (a) & (b) & (c) \\
          \includegraphics[clip,width=0.30\textwidth]
          {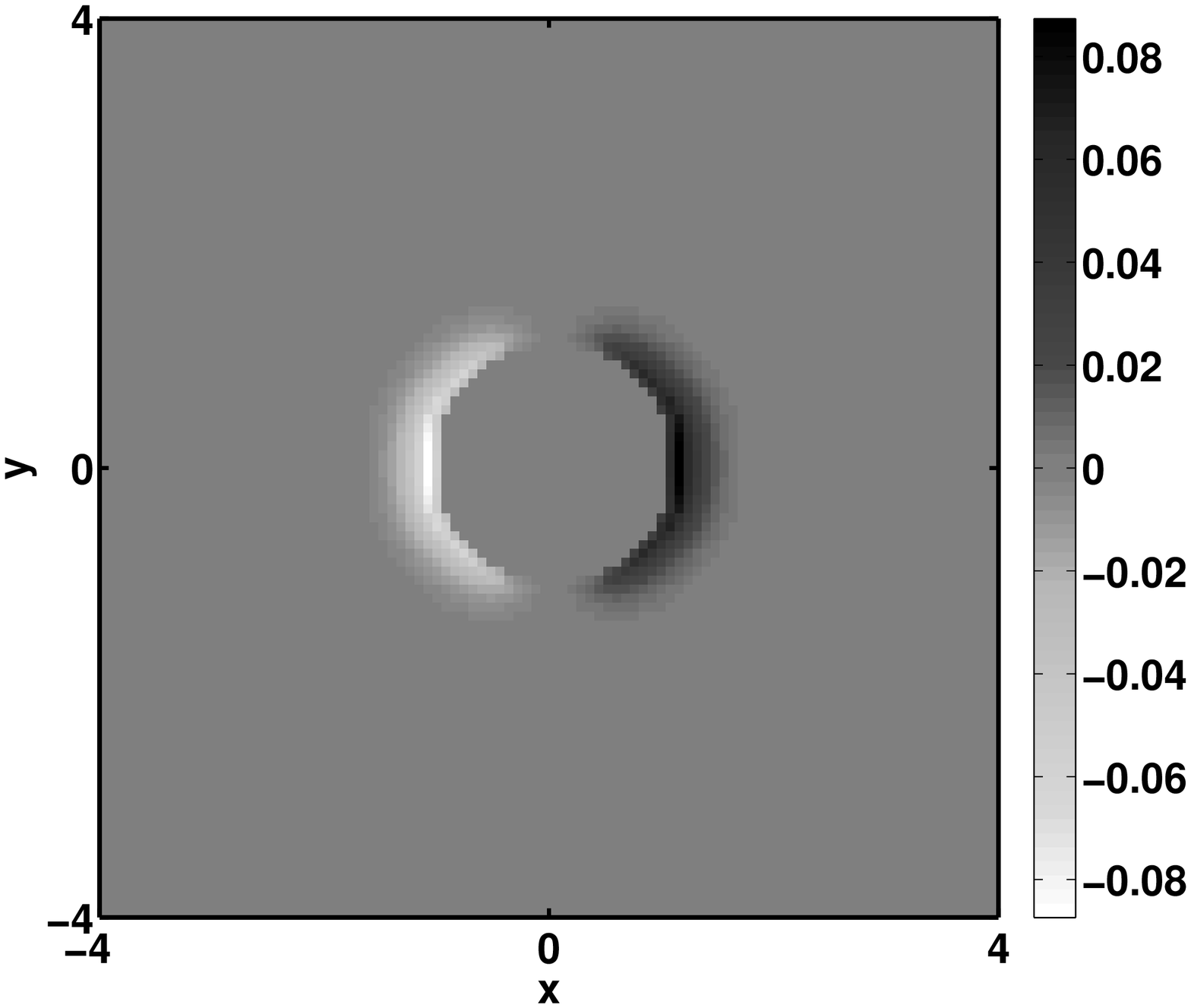}
      & 
          \includegraphics[clip,width=0.30\textwidth]
          {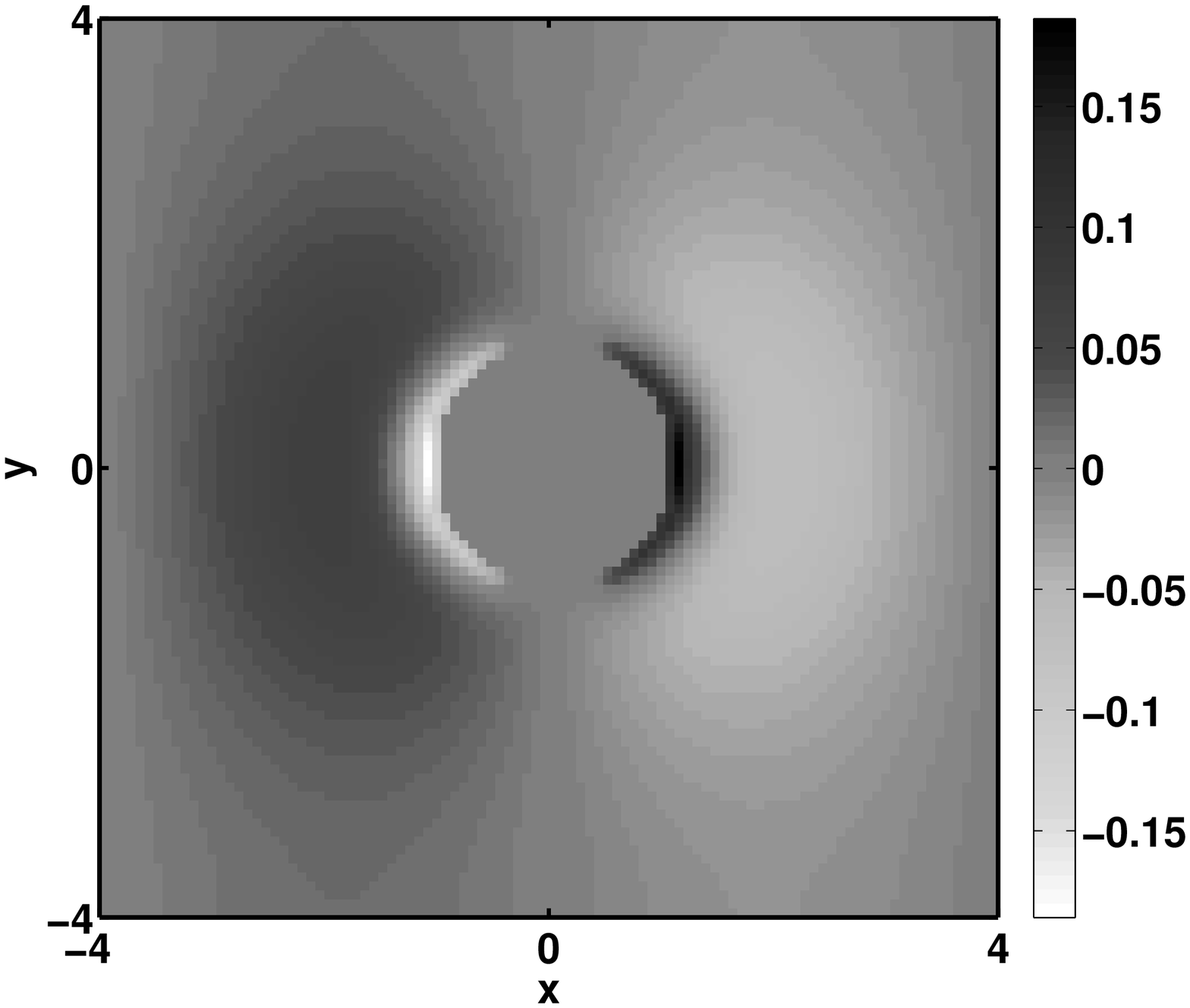}
      & 
          \includegraphics[clip,width=0.30\textwidth]
          {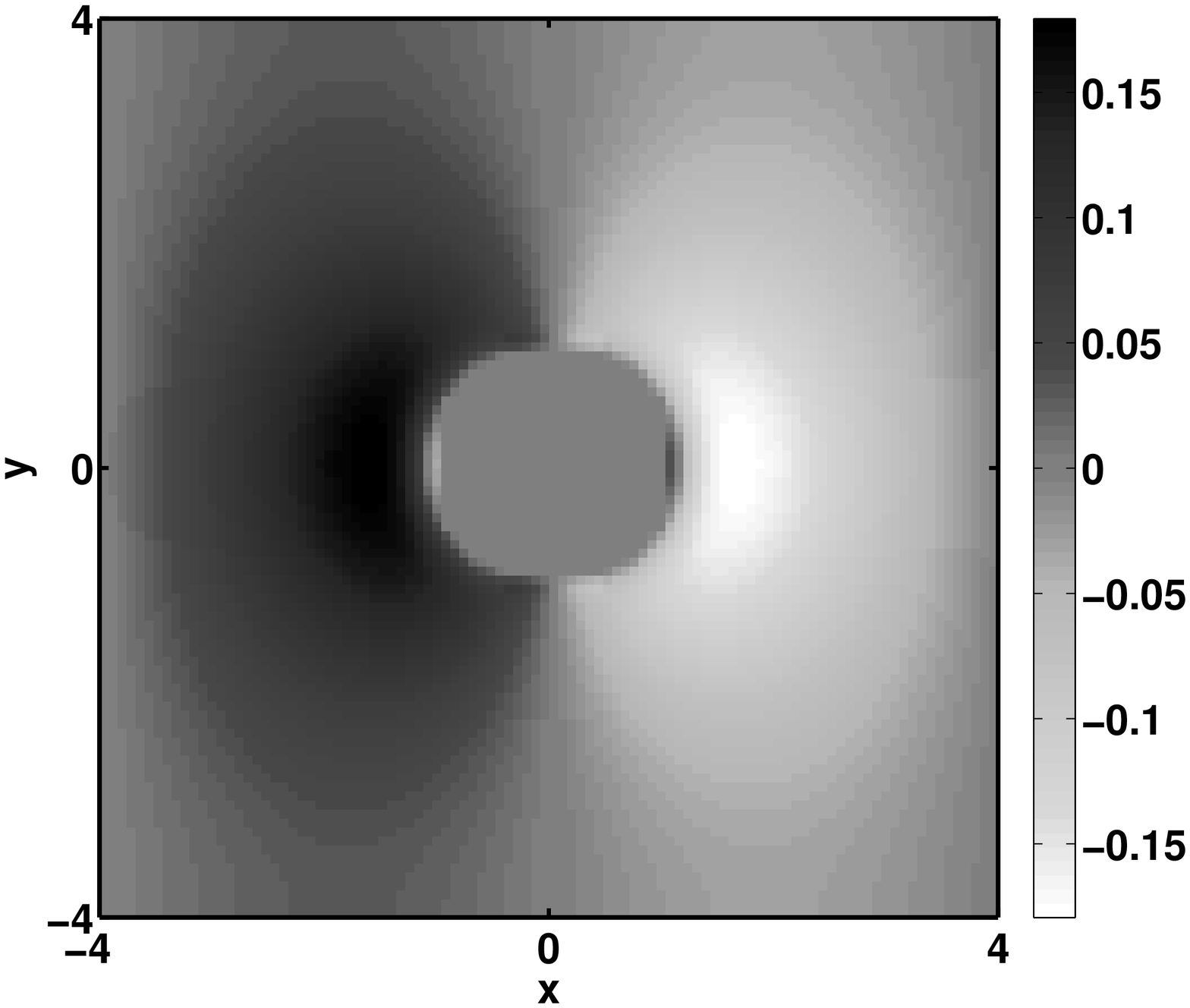}
      \end{tabular}
    \end{center}
    \caption{$2$--dimensional cuts of the first--order concentration
      profile of the negative salt ions ${\tilde c}_{-}^{(1)}$, for
      the parameters $d=0.08$, $\tilde R=1$, $\Phi\simeq8.18\cdot
      10^{-3}$ and an electric field of ${\tilde E}_x=1$ acting in the
      $x$ direction. The reduced charges of the sphere and the dipole
      moments of the system are (a) $\hat Z=0.0$, ${\tilde
        p}_x=-6.44$, (b) $\hat Z=1.8$, ${\tilde p}_x=0.087$ and (c)
      $\hat Z=3.0$, ${\tilde p}_x=11.16$.}
    \label{fig: lowcharge cminus 0.08}
  \end{figure}

  \begin{figure}
    \begin{center}
      \begin{tabular}{c c c}
        (a) & (b) & (c) \\
           \includegraphics[clip,width=0.30\textwidth]
           {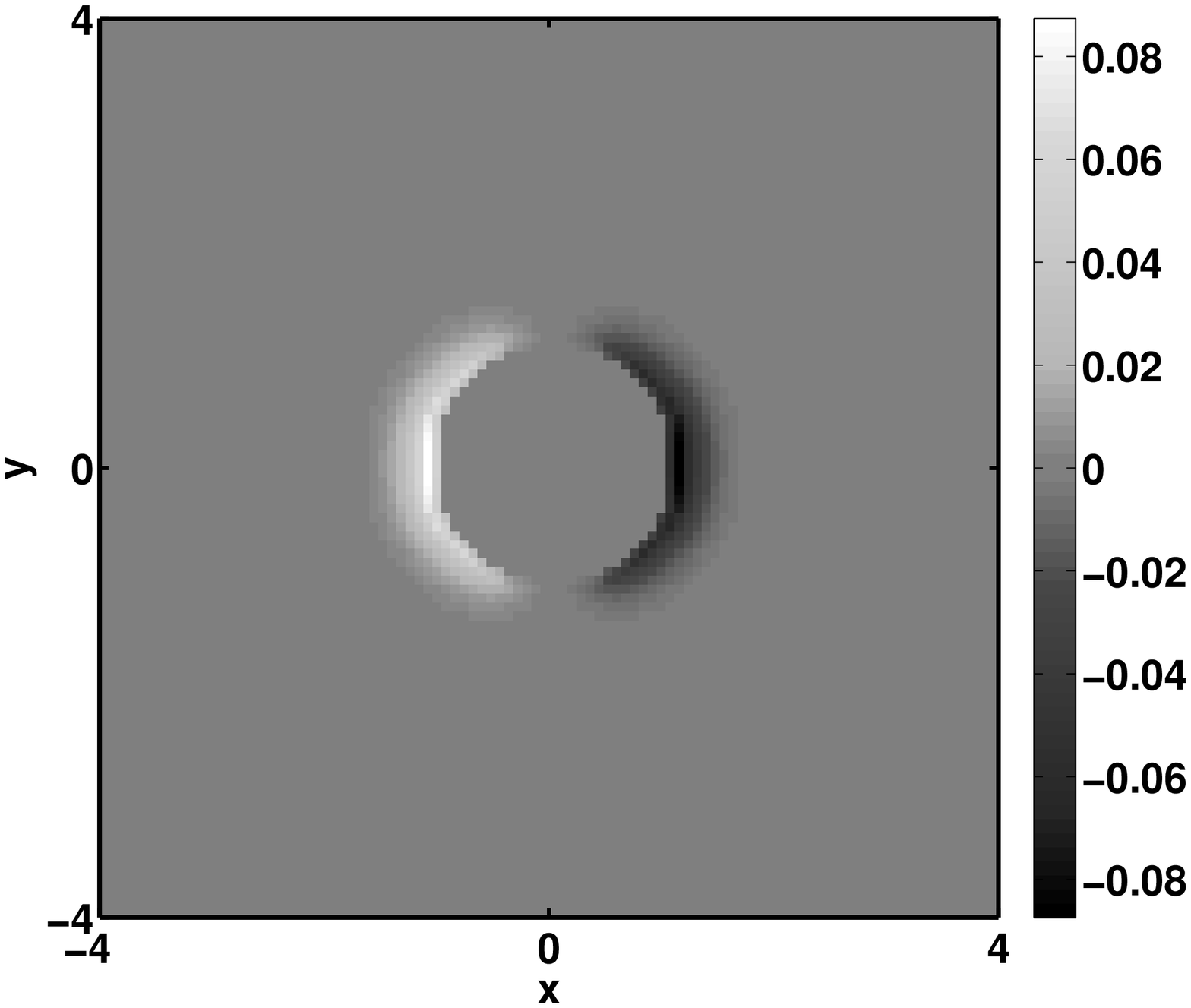}
      & 
           \includegraphics[clip,width=0.30\textwidth]
           {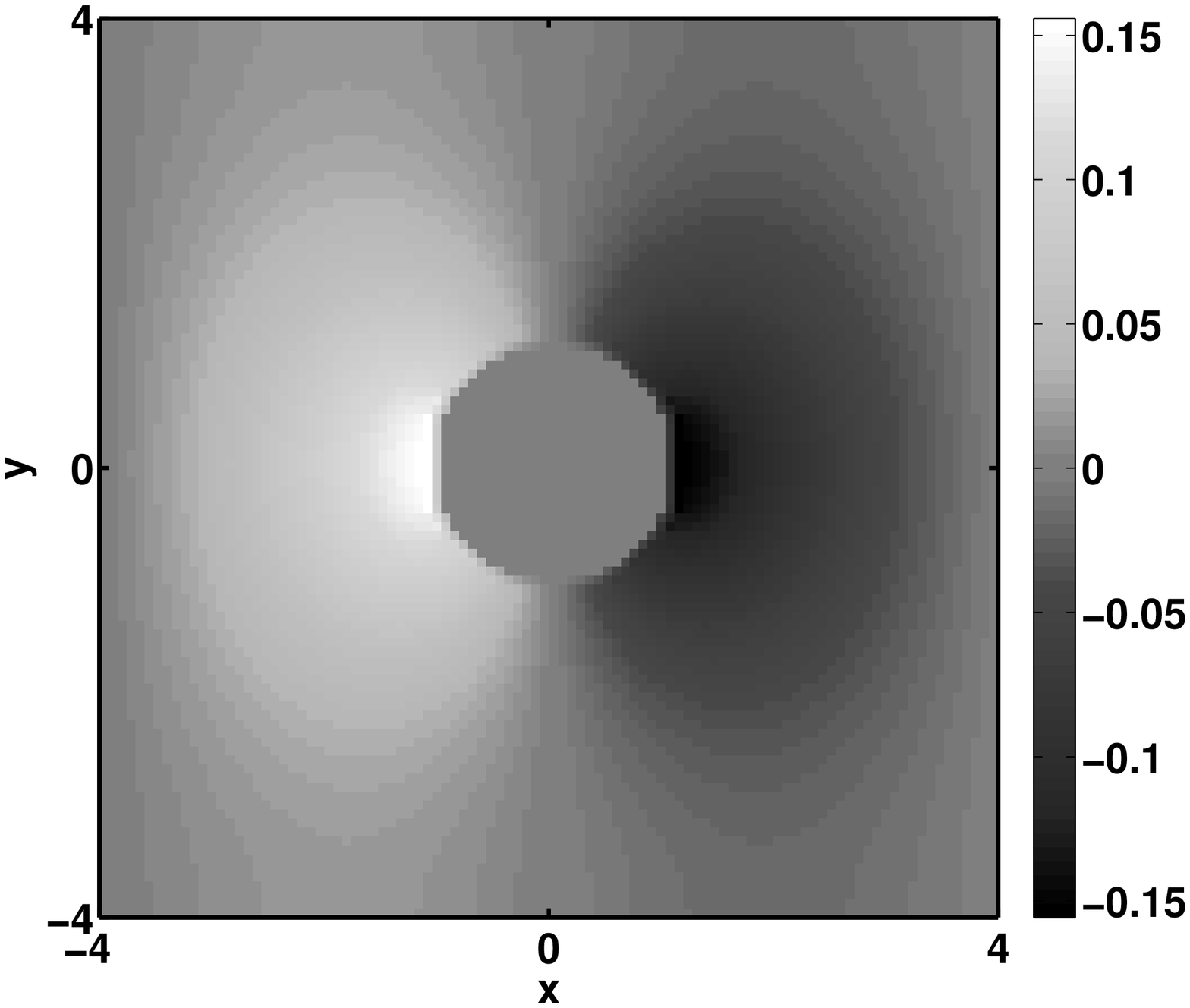}
      & 
           \includegraphics[clip,width=0.30\textwidth]
        {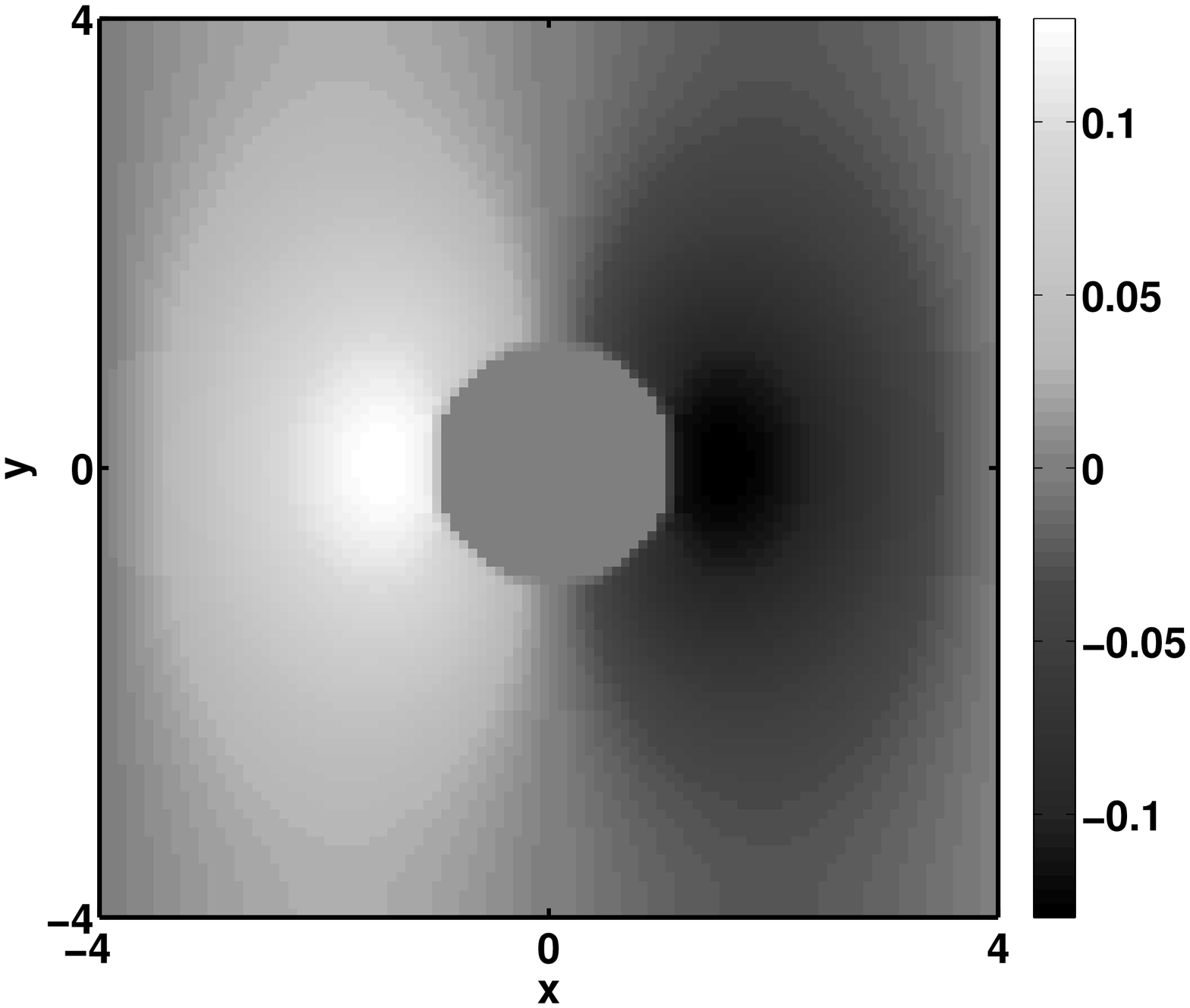}
      \end{tabular}
    \end{center}
    \caption{$2$--dimensional cuts of the positively charged salt
      concentration ${\tilde c}_{+}^{(1)}$, for the same sets of
      parameters as in Fig. \ref{fig: lowcharge cminus 0.08}.}
    \label{fig: lowcharge cplus 0.08}
  \end{figure}

  \begin{figure}
    \begin{center}
      \begin{tabular}{c c c}
        (a) & (b) & (c) \\
           \includegraphics[clip,width=0.30\textwidth]
           {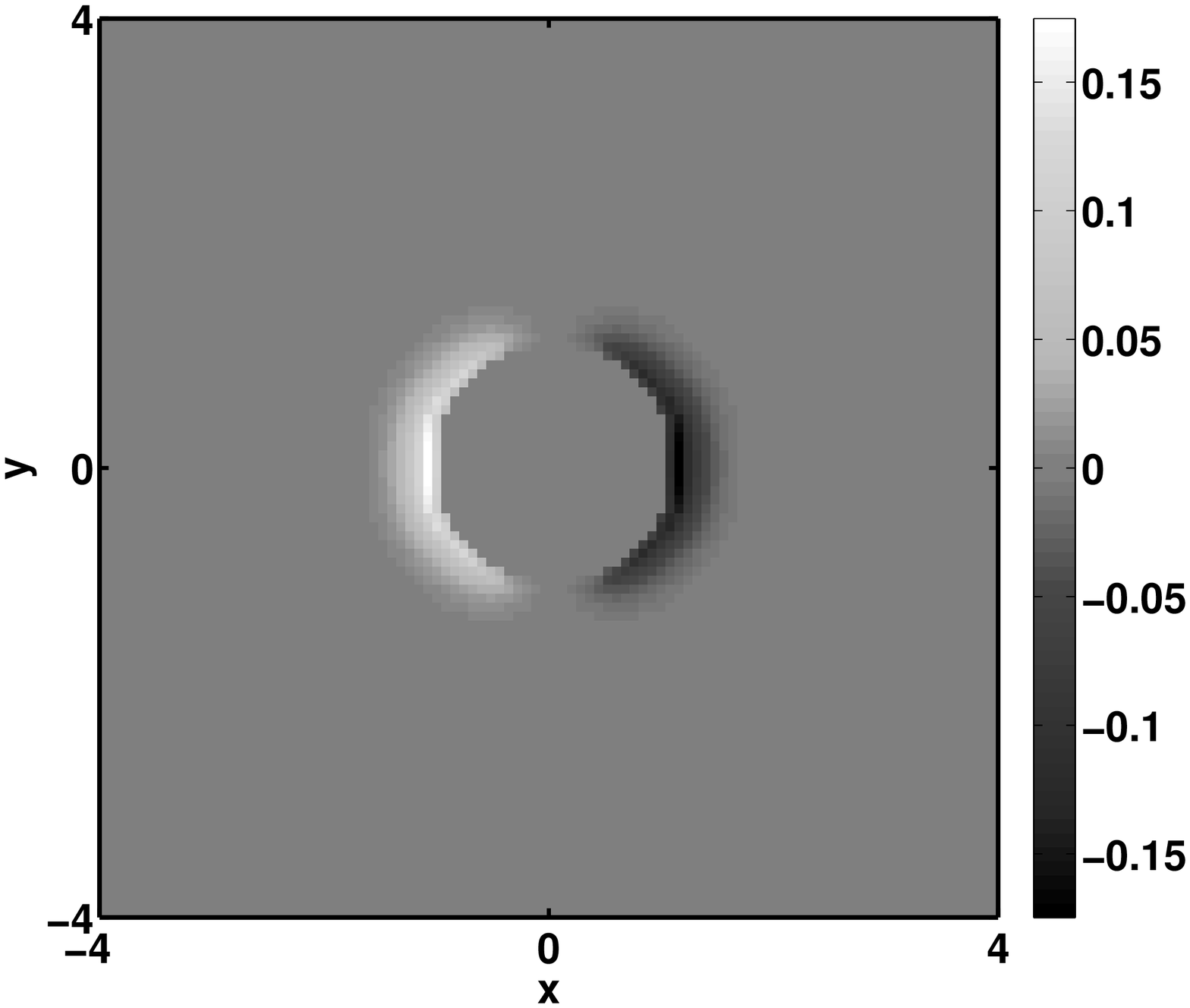}
      & 
           \includegraphics[clip,width=0.30\textwidth]
           {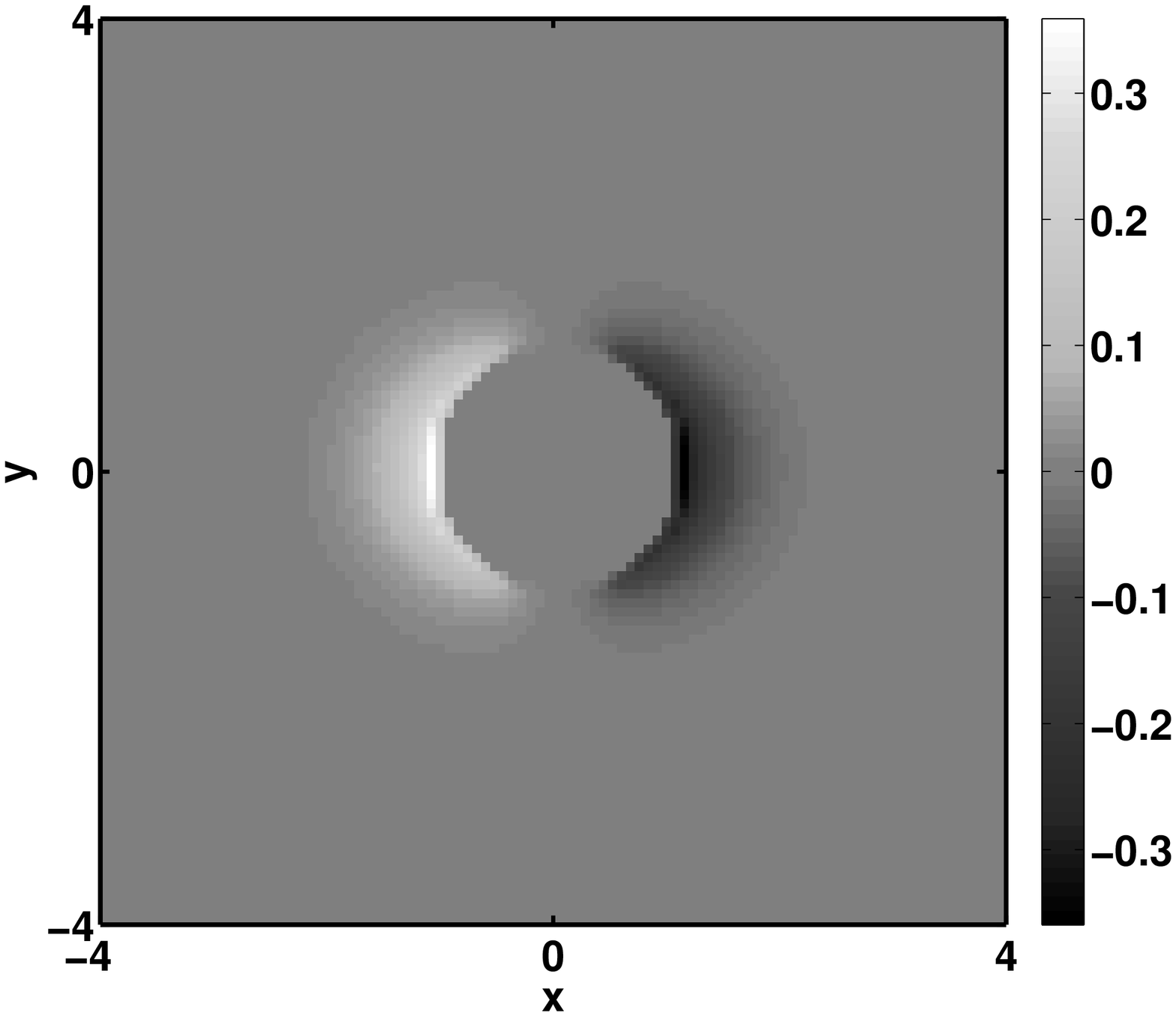}
      & 
           \includegraphics[clip,width=0.30\textwidth]
           {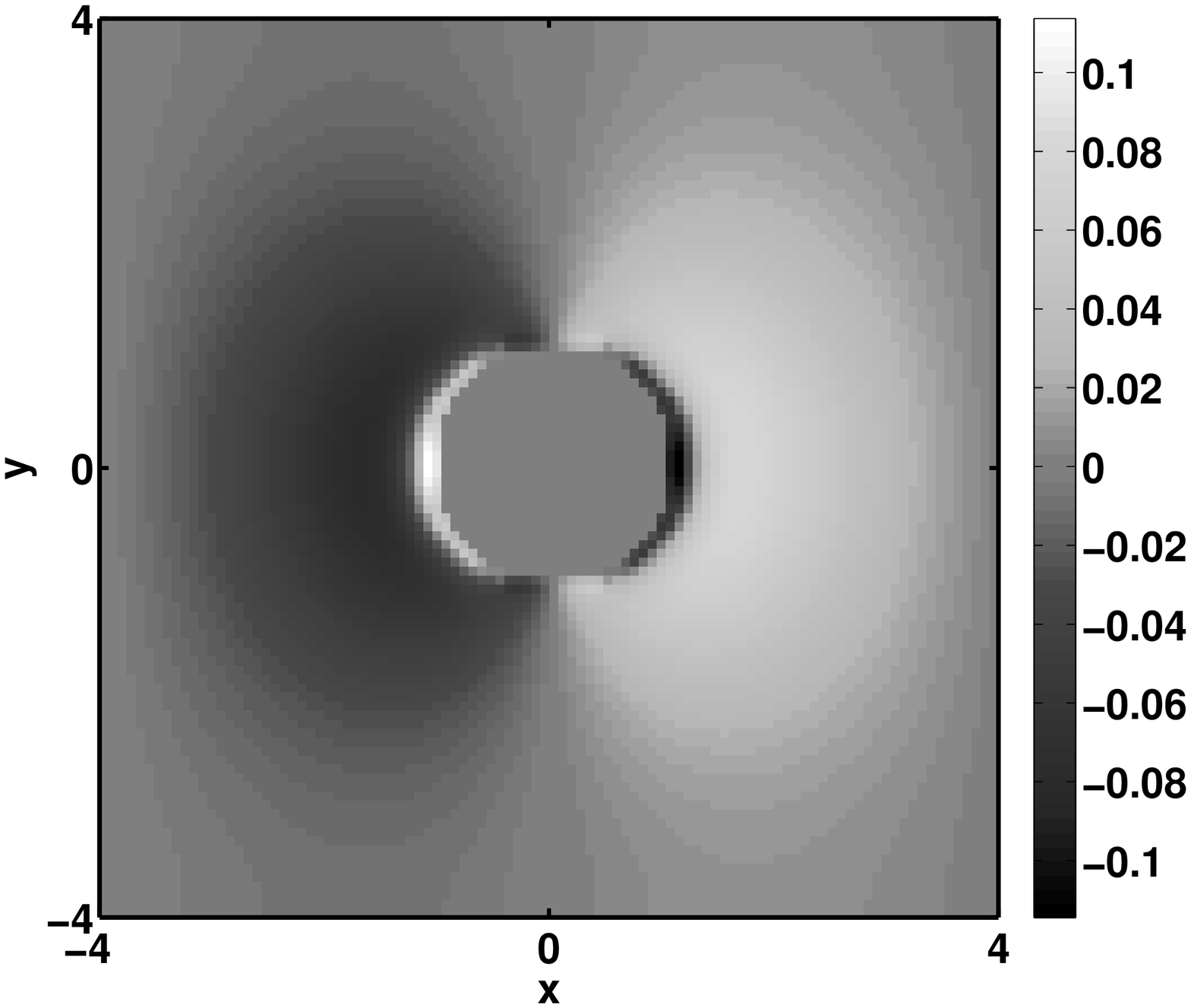}
      \end{tabular}
    \end{center}
    \caption{$2$--dimensional cuts of the first order charge density
      ${\tilde \rho}^{(1)}$. The parameters are the same as in the
      previous images (see Fig. \ref{fig: lowcharge cminus 0.08}).}
    \label{fig: lowcharge rho 0.08}
  \end{figure}

  \begin{figure}
    \begin{center}
      \includegraphics[clip,width=0.5\textwidth]
      {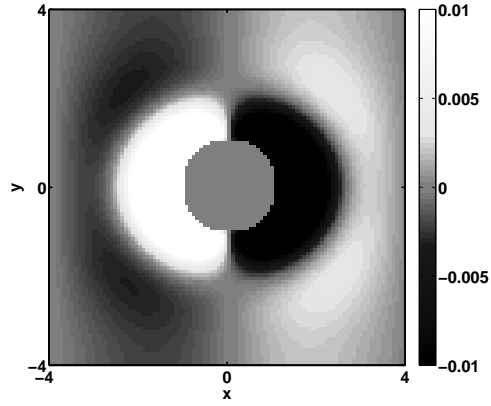}
    \end{center}
    \caption{Same data as Fig. \ref{fig: lowcharge rho 0.08}(b)
      but with rescaled colormap.}
    \label{fig: lowcharge rho 0.08 Z 1.8}
  \end{figure}

  Returning to Fig. \ref{fig: dipole vs charge}, we observe that the
  dipole moment increases with increasing charge. For a reduced charge
  of about $\hat Z = 2$, depending on the volume fraction, the sign of
  the dipole moment changes, i.~e. the orientation of the charge cloud
  is reversed. In order to elucidate the phenomenon in some more
  detail, we present in Figs. \ref{fig: lowcharge cminus 0.08},
  \ref{fig: lowcharge cplus 0.08} and \ref{fig: lowcharge rho 0.08}
  the first--order charge clouds as two--dimensional cuts in the $xy$
  plane (the field is oriented in $x$ direction): Figure \ref{fig:
    lowcharge cminus 0.08} depicts the negative salt ions,
  Fig. \ref{fig: lowcharge cplus 0.08} the positive salt ions and
  Fig. \ref{fig: lowcharge rho 0.08} the charge density. Most
  interesting are the figures in the case of a very small but already
  ``normal'' dipole moment: Here one sees that the orientation of the
  charge cloud near the colloid is still ``anomalous'', but this is
  more than compensated from ``normal'' contributions further
  away. For the charge distribution, this cut is plotted again in
  Fig. \ref{fig: lowcharge rho 0.08 Z 1.8} with a rescaled colormap
  for better visibility. All in all, this charge cloud reversal
  highlights that in electrophoresis both electrostatic and
  hydrodynamic effects are important, and that these may compete,
  resulting in a change of qualitative behaviour depending on
  conditions.

\begin{figure}
    \begin{center}
      \includegraphics[clip,width=0.7\textwidth]
      {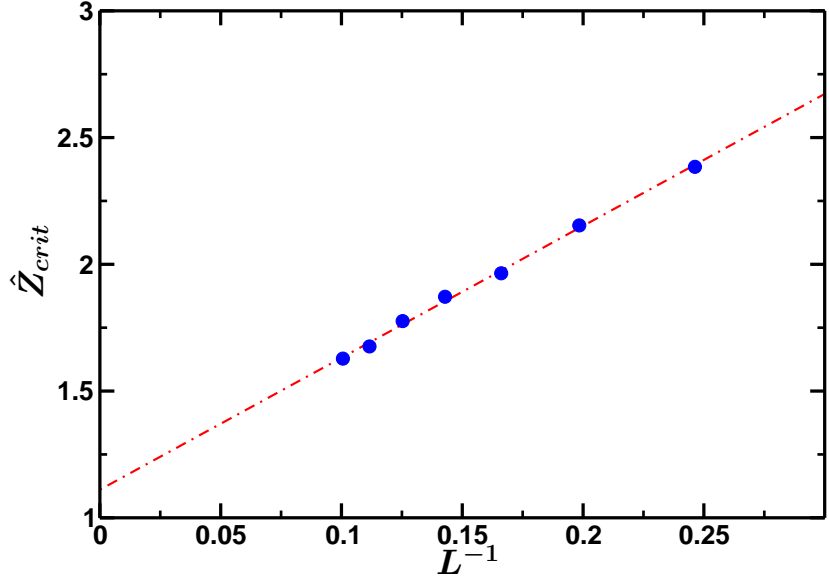}
    \end{center}
    \caption{Critical charge as function of the inverse box length
      $L^{-1}$ as obtained from the roots of spline functions.  The
      line is a linear fit to the data points.}
    \label{fig:critical_charge}
  \end{figure}

  The critical charge, i.~e. the value of the colloid charge at which
  the dipole moment switches its sign, depends on the volume fraction.
  Keeping all other parameters fixed as before and the resolution at
  $d = 0.07$, we mapped out the functions $\tilde p$ vs. $\hat Z$ and
  determined the critical value via spline interpolation. The
  dependence of this value on the linear system size $L$ is shown in
  Fig. \ref{fig:critical_charge}; extrapolation yields
  \begin{equation}
      {\hat Z}_{crit} (\Phi \to 0) = 1.11 \pm 0.02 \,.
  \end{equation}

  \begin{figure}
    \begin{center}
      \includegraphics[clip,width=0.70\textwidth]
      {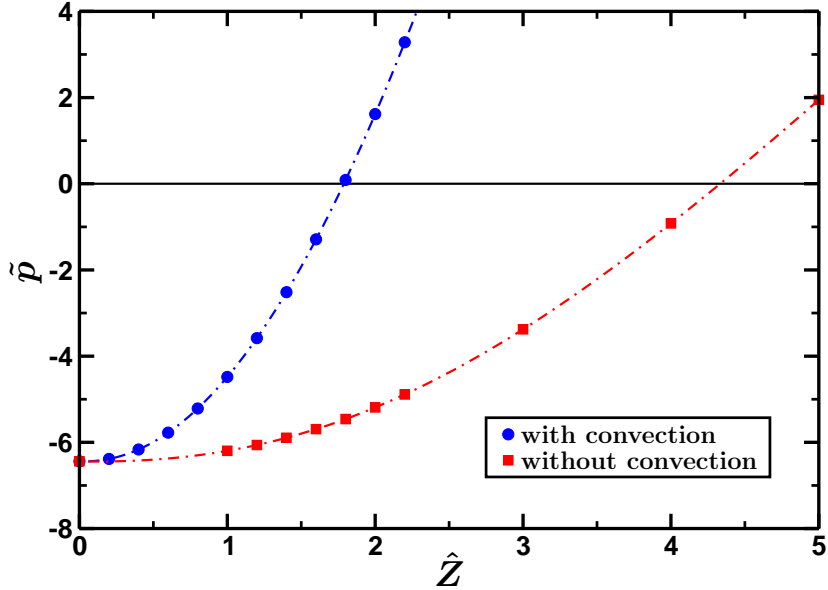}
    \end{center}
    \caption{Dipole moment as function of the reduced charge for the
      same parameters as in Fig. \ref{fig: dipole vs charge}.  Here,
      we choose $d=0.08$ and $\Phi\simeq8.18\cdot10^{-3}$. The blue
      circles are the same data as before. For the red squares the
      convection velocity was set to $\tilde{\boldsymbol{v}}^{(1)}
      \equiv 0$.  The dash--dotted lines are cubic splines.}
    \label{fig: P vs Z without convection}
  \end{figure}

  Finally, Fig. \ref{fig: P vs Z without convection} demonstrates that
  the charge cloud reversal is fairly strongly affected by the value
  of the ion diffusion coefficient. To this end, we also studied the
  case $D_i \to \infty$, which we realized computationally by just
  repeating our calculation with $\tilde D = 1.5$ but turning the
  convection term in the convection--diffusion equation
  off. Alternatively, one may therefore view this calculation as a
  study that elucidates the influence of convection. The result
  is clearly that convective transport helps in establishing the
  ``normal'' orientation.

  \section{Concluding remarks}
  \label{sec:conclusions}

  In this paper we investigated a new numerical approach for the
  theoretical treatment of a charge--stabilised colloidal dispersion
  in an external electric field. The system, given by a solid charged
  sphere in an electrolyte solution, was treated on a Mean--Field
  level, resulting in a system of coupled nonlinear partial
  differential equations. Following the ideas of O'Brien and White
  \cite{OBr78}, the nonlinearity is confined to the equilibrium
  Poisson--Boltzmann equation by application of a linearisation with
  respect to the external field. The iterative procedure in
  combination with the chosen specialised solvers turns out to be very
  efficient and only limited by the choice of the lattice
  spacing. While the demand on memory for the bulk methods increases
  linearly with the number of grid nodes, the surface integral solver
  for the Stokes equation requires a dense matrix connecting all
  surface nodes of the colloidal particles. Thus, the amount of memory
  needed for the storage of this matrix increases rapidly with the
  resolution of the sphere. Therefore this method is very efficient up
  to a certain value of the resolution; beyond that, alternative
  solutions must be developed. However, the iterative method has the
  advantage that it is designed as a modular solver and every module
  can be replaced via an alternative algorithm. One possible approach
  for going to higher resolutions is to replace the Stokes solver by a
  bulk method. For example, the time step of a lattice Boltzmann
  method could be adjusted such that it is identical with the time
  step of the convection--diffusion solver. Thus, the iterative method
  could be modified in a way that the Nernst--Planck and the Stokes
  equation are solved simultaneously. Furthermore, the lattice
  Boltzmann method would have the same degree of locality as the
  convection--diffusion equation solver, and hence parallelisation
  using a domain decomposition would be easily
  implemented. Nevertheless, the current single--processor
  implementation is quite efficient and reliable for a fairly
  satisfactory range of parameters. The numerical results agree
  reasonably well with various established results from the
  literature.

  Furthermore, all parameters in the method are controlled
  independently, which offered, e.~g., the opportunity to study the
  dependence of the electrophoretic mobility on the diffusion
  coefficient of the surrounding ions. One of the most interesting
  results presented above is the conclusion that the screening
  mechanism has an effect on the electrophoretic mobility, i.~e. the
  mobility varies by a few percent if the amount of salt is increased
  in the solution, while the screening length is kept constant. This
  shows clearly that the assumption that counterion--dominated systems
  may be mapped onto salt--dominated systems is only approximately
  true: If the accuracy goes beyond, say, $5 \%$, special care must be
  taken for the exact screening mechanism. Moreover, this dependence
  is also affected qualitatively by the diffusion coefficient.

  Another very interesting application is the examination of weakly
  charged colloidal systems. If an electric field acts on an uncharged
  colloidal sphere in salt solution, ions move and are deflected at
  the surface of the solid particle, resulting in an ``anomalous''
  dipole moment anti--parallel to the driving field. This
  ``anomalous'' dipole moment was recently addressed by Dhont and Kang
  \cite{Dhont2010} analytically. With our numerical method we were
  able to reproduce their result up to one percent
  difference. Increasing the colloid charge, we find a critical value
  at which the dipole moment changes its sign and the ion cloud
  reverses its orientation.

  All in all, the developed tool is computationally much cheaper than
  the raspberry MD/LB model \cite{Lob2004}, while having a somewhat
  broader range of applicability than the original work of O'Brien and
  White \cite{OBr78}. Clearly, it has limitations, as outlined in more
  detail in the Introduction. Although it will therefore not be able
  to study all electrokinetic phenomena in charge--stabilised
  colloidal dispersions --- in particular, many--colloid systems where
  the particles continuously move with respect to one another, and a
  global rest frame does not exist, are out of reach for the present
  single--colloid version --- we believe that it has already proven
  useful and is fairly likely to continue to do so.

  \section*{Acknowledgments}

  This work was funded by the SFB TR 6 of the Deutsche
  Forschungsgemeinschaft. We thank J. K. G. Dhont, B. Li and
  A. J. C. Ladd for helpful discussions.

  \section*{References}

  \bibliographystyle{unsrt}
  \bibliography{eke}

\end{document}